\documentclass[referee,useAMS,usenatbib]{mn2e}
\usepackage{graphicx,graphics,amsmath,amssymb,mathrsfs}

\newcommand{\nin}{\noindent}
\newcommand{\dmath}{\mbox{${\mathrm d}$}}

\newcommand{\der}[2]{\frac{\dmath{#1}}{\dmath{#2}}}
\newcommand{\DER}[2]{\frac{\partial{#1}}{\partial{#2}}}
\newcommand{\dder}[2]{\frac{\dmath\,^2{#1}}{\dmath{#2}^2}}

\newcommand{\pom}{\dot{\varpi}}
\newcommand{\mscr}[1]{\displaystyle{\mathscr {#1}}}

\newcommand{\siga}[2]{\Sigma_{#1}^{#2}}
\newcommand{\bfpar}[3]{{\bf{#1}}_{#2}^{#3}}
\newcommand{\vtilde}{{\tilde{v}}_{\phi}}

\newcommand{\rmi}{{\rm i}}

\newcommand{\rme}{{\rm e}}

\newcommand{\intg}[1]{\int_{-\infty}^{#1}}
\newcommand{\intglwr}[1]{\int^{\infty}_{#1}}
\newcommand{\intgbth}[2]{\int_{#1}^{#2}}
\newcommand{\intgl}{\int_{-\infty}^{\infty}}

\newcommand{\rmpar}[1]{{\rm {#1}}}
\newcommand{\sigd}[1]{\Sigma_{\rmpar{d}}^{#1}}
\newcommand{\md}{M_{\rmpar{d}}}

\def\apj{{Astroph.\@ J. }}
\def\mnras{{Mon.\@ Not.\@ Roy.\@ Ast.\@ Soc.}}

\def\aj{{Astron.\@ J.}}

\def\apjl{{Astrophysical.\@ J. {\rm Letters}}}

\begin{document}

\title[Modal theory of collisionless discs]
{A modified WKB formulation for linear eigenmodes of a collisionless 
self-gravitating disc in the epicyclic approximation}

\author[Gulati \& Saini]{Mamta Gulati$^{1,2,3,4}$, Tarun Deep Saini$^{3,   
  5}$ \\
  $^{1}$ Indian Institute of Science Education and Research Mohali, Chandigarh, 560 012, India\\
  $^{2}$ Raman Research Institute, Sadashivanagar, Bangalore 560 080, India\\
  $^{3}$ Indian Institute of Science, Bangalore 560 012, India\\	
  $^{4}$ mgulati@iisermohali.ac.in\,,
  $^{5}$ tarun@physics.iisc.ernet.in\,,	
  }
\maketitle

\begin{abstract}
The short--wave asymptotics (WKB) of spiral density waves in self-gravitating
stellar discs is well suited for the study of the dynamics of tightly--wound 
wavepackets. But the textbook WKB theory is not well adapted to the study of 
the linear eigenmodes in a collisionless self-gravitating disc because of 
the transcendental nature of the dispersion relation. We present a modified WKB 
of spiral density waves, for collisionless discs 
in the epicyclic limit, in which the perturbed gravitational potential is related to the 
perturbed surface density by the Poisson integral in Kalnaj's logarithmic spiral form.  
An integral equation is obtained for the surface density perturbation, which is 
seen to also reduce to the  standard WKB dispersion relation. We specialize to
a low mass (or Keplerian) self-gravitating disc around a massive black hole, 
and derive an integral equation governing the eigenspectra and eigenfunctions 
of slow precessional modes. For a prograde disc, the integral kernel turns out be 
real and symmetric, implying that all slow modes are stable. We apply the slow mode
integral equation to two unperturbed disc profiles, the Jalali--Tremaine 
annular discs, and the Kuzmin disc. We determine eigenvalues and eigenfunctions for both $m = 1$ 
and $m = 2$ slow modes for these profiles and discuss their properties. Our results compare well 
with those of Jalali--Tremaine. 
\end{abstract}


\section{Introduction}

\nin
Astrophysical discs display rich structural features such as: double-peak distribution of light 
in the central regions of galaxies like NGC$4486$B (elliptical galaxy), and M$31$ (spiral galaxy) 
\citep{lau93, lau96}; lopsided brightness distribution of scattered light, warp and clumps in the 
disc around $\beta$ Pictoris, which is the second brightest star in the constellation Pictor 
\citep{heap00,tel00}; spiral structure in HD $141569$A \citep{clamp03}; clumpy rings in Vega 
\citep{marsh06}, and many more. Structural and kinematic properties of several astrophysical 
systems have been found to be correlated to the global properties of the system.  For example, 
over a sample of thousands of galaxies, a correlation has been found between the lopsidedness, 
Black Hole (BH) growth, and the presence of young stellar populations in the center of a galaxy 
\citep{richrd09}. Therefore, the dynamics of these disc-like systems is an important field of 
investigation.
 
In this work we will mainly address discs around massive central objects where the self-gravity 
of the disc is sufficiently weak to make the disc nearly Keplerian. Such discs can support 
eigenmodes that are slow in comparison to the Keplerian flow. The discs could be gaseous, or 
particulate and thus collisionless. \cite{st10} \& \cite{gss12} studied nearly Keplerian discs 
by treating the disc as fluid. By assuming that the disc particles interact through softened-gravity, 
they could mimic the behaviour of collisionless discs. A major limitation of this method is that 
the fluid discs do not support slow modes for azimuthal wavenumber number $m > 1$. Real 
collisionless discs, however, can support modes with all values of $m$ \citep{jt12}. Softened 
gravity discs have also been studied by \citet{tre01}, but with subtle differences in method, 
as discussed in detail in \citet{gss12}.

A general eigenvalue formulation exists for fluid discs, even without assuming the slow mode 
approximation \citep{gold79}, although this formulation assumes the tight-winding approximation 
to make the gravitational potential due to the disc perturbations local. Recently, simulations 
involving linear perturbations in nearly Keplerian collisionless discs have been performed by 
\citet{jt12} using the finite element method. However, apart from the disc stability analysis, 
which gives useful but limited analytical treatment of collisionless discs. In literature there 
does not exist an eigenvalue formulation for collisionless discs in the WKB approximation \citep{bt08}. 
In this paper we propose an \emph{eigenvalue formulation of linear perturbations in a collisionless 
disc} based on the assumptions:
\begin{enumerate}
\item Radial wavenumber $k$ times the radius $R$ is much larger than the azimuthal 
wavenumber $m$, i.e. $|kR| \gg m$.
\item The epicyclic approximation, in which the velocity dispersion 
$\sigma_R$ is much less than the circular speed $v_c$. We also assume 
that 
\begin{equation*}
\frac{\sigma_R}{v_c} \sim \frac{m}{|kR|} \ll 1 \,. 
\end{equation*}
\end{enumerate}
\nin
The  WKB or the tightly-wound spiral approximation requires $|kR| \gg m$, although, 
in many cases the approximation works fairly well even for $|kR|$ as small as unity 
\citep{tre01,gss12, jt12}. The gravitational potential for a tightly-wound spiral is 
local, that is, it can be obtained from the local perturbed density \citep{bt08}. 
However, in our formulation we do not require this assumption, since we use the 
logarithmic-spiral decomposition of both the perturbed surface density and the 
gravitational potential \citep{klnjs71} for solving the Poisson's equation. This allows 
the potential to take contributions from perturbations across the disc. The integral 
equation is derived without restricting ourselves to Keplerian discs, and could be 
used to explore eigenmodes of non-Keplerian discs such as galactic disc. However, 
to explore the validity of our formulation, for this work we restrict ourselves to 
nearly Keplerian discs and further approximate the integral equation to the slow 
mode case to make comparisons with the results of \citet{jt12}. As a further test 
case we also consider perturbations in a Kuzmin disc.

In \S~\ref{unprtbd_disc} we describe the unperturbed disc. The integral equation is 
derived in \S~\ref{int_eqn}. We show in the Appendix~\ref{wkbreduction} that the integral 
equation reduces to the standard WKB dispersion relation under the local approximation. 
We take the {\it slow mode} limit of the integral equation in \S~\ref{slwmode} which reduces 
the equation to a linear eigenvalue problem. In \S~\ref{num_mthd} we describe the 
numerical method adopted to solve the eigenvalue problem. \S~\ref{num_sol} gives an 
account of solutions obtained for two surface density profiles (1) JT annular disc, the 
disc considered in \citet{jt12} (2) Kuzmin disc. We conclude with general remarks in 
\S~\ref{cnclsn}.

\section{Unperturbed disc}\label{unprtbd_disc}

Dynamics of disc around a massive compact object, like discs of stars 
orbiting the super-massive BH found at the center of most galaxies, 
debris discs etc, are governed by the gravitational potential of the 
central mass and the self gravity of the disc. In such situations 
usually the ratio of disc mass $\md$ to central mass $M$ (i.e. 
$\varepsilon \equiv \md/M$) is much smaller than unity. Orbits of disc 
particles are nearly Keplerian, and since we neglect relativistic 
effects the discussion is applicable several Schwarzschild radius 
away from the central BH. Henceforth we proceed by approximating our 
disc to be razor-thin, i.e. we restrict ourselves to $z = 0$ plane and work 
in two spatial dimensions ($\bfpar{r}{}{} \equiv (R,\phi)$, cylindrical 
polar coordinates are used here) .

The potential $\Phi_0(R)$ for the unperturbed disc is given by,
\begin{equation}
 \Phi_0(R) \;=\; -\frac{G M}{R} + \Phi_{\rm d}(R)\,,\label{phi}
\end{equation}
\nin 
which is a sum of the Keplerian potential due to central mass and
gravitational potential due to self gravity of the disc:

\begin{equation}
 \Phi_{\rm d}(\bfpar{r}{}{}) \;=\; -G\int \frac{\sigd{}(\bfpar{r'}{}{})}
{|\bfpar{r}{}{} - \bfpar{r'}{}{}|} \dmath^2r'\,.\label{phid}
\end{equation}

\nin 
Since disc mass is $\mathnormal{O}(\varepsilon)$ smaller compared to central mass, 
so is $\Phi_{\rmpar{d}}$ compared to Keplerian potential. Nearly circular orbits 
have radial frequency $\kappa$ and azimuthal frequency $\Omega$ given by
\begin{align}
 \Omega^2(R) & \;=\; \frac{GM}{R^3} + \frac{1}{R}\der{\Phi_{\rm d}}{R}\,,\label{omega}\\
 \kappa^2(R) & \;=\; \frac{GM}{R^3} + \frac{3}{R}\der{\Phi_{\rm d}}{R} + \dder{\Phi_{\rm d}}{R}\,.\label{kappa}
\end{align}
\nin 
Such nearly circular orbit precess at a rate given by 
\begin{align}
 \pom(R) & \;=\; \Omega(R) - \kappa(R)\nonumber\\
& \;=\; -\frac{1}{2\Omega(R)}\left(\frac{2}{R}\der{}{R} + \dder{}{R}\right)\Phi_{\rmpar{d}}(R) 
+ \mathnormal{O}(\varepsilon^2)\,.\label{pom}
\end{align}
\nin 
The unperturbed stellar orbits are considered to be nearly circular. 
The phase-space coordinates of these epicyclic orbits are given by \citep{bt08};
\begin{align}
R' &\,=\, R + \frac{\gamma\vtilde}{\kappa}\left(1 - \cos(\tau)\right) + \frac{v_R}{\kappa}\sin(\tau)\,,\nonumber\\
\phi' \,=\, \phi + \frac{\Omega \tau}{\kappa} &+ \frac{\gamma\gamma'}{2\kappa}\vtilde\tau + 
\frac{\gamma}{R\kappa}\left[\gamma\vtilde\sin(\tau) - v_R\left(1 - \cos(\tau)\right)\right]\,,
\label{epi_approx1}
\end{align}
and
\begin{align}
v_R' &\,=\, v_R\cos(\tau) + \gamma\vtilde\sin(\tau)\,,\nonumber\\
\gamma\vtilde ' &\,=\, \gamma\vtilde\cos(\tau) - v_R\sin(\tau)\,.
\label{epi_approx2}
\end{align} 
\nin
where $\tau \,=\, \kappa_g(t' - t)$;\, $\kappa_g \;=\; \kappa(R_g)$;\, $R_g$ 
is the mean radius of the orbit for a given angular momentum; and 
$\gamma'$ is the derivative of $\gamma(R) \;=\; 2\Omega(R)/ \kappa(R)$ 
w.r.t. $R$. Also, $\vtilde(R) = v_{\phi}(R) - v_c(R)$. At $\tau = 0$, 
the phase-space coordinates $(\bfpar{r'}{}{},\bfpar{v'}{}{}) = (\bfpar{r}{}{},\bfpar{v}{}{})$.
 
The phase space distribution function for the unperturbed collisionless disc in 
the epicyclic approximation is given by the Schwarzschild distribution function (DF)
\begin{equation}
 f_0(R,v_R,\vtilde) \;=\; \frac{\gamma \sigd{}(R)}{2\pi\sigma_R^2}\exp
 \left(-\,\frac{v_R^2 + \gamma^2\vtilde^2}{2\sigma_R^2}\right)\,,\label{sch2D}
\end{equation}
\nin
where $\sigd{}(R)$ is the unperturbed surface density profile and 
$\sigma_R$ is the radial component of the velocity dispersion.

\section{Perturbed disc}\label{int_eqn}
To study the evolution of small perturbations in this system we begin with 
perturbing the initial DF such that the distribution function at any time, 
$t$ is given by,
\begin{equation}
f(R,\phi,v_R,{\tilde{v}}_{\phi},t) = f_0(R,v_R,{\tilde{v}}_{\phi}) + f_1(R,\phi,v_R,{\tilde{v}}_{\phi},t)\,. 
\label{f_per}
\end{equation}
where $f_0$ is the unperturbed DF, and the perturbation 
$f_1 \sim \varepsilon f_0$. Hereafter, all perturbed quantities are denoted 
with a subscript $1$ such as $X_1$. Perturbations in the surface density are related to the perturbed DF as
\begin{equation}
\siga{1}{}(R,\phi,t) = \int f_1(R,\phi,v_R,{\tilde{v}}_{\phi},t) \dmath^2\bfpar{v}{}{}\,,\label{den_per}
\end{equation}
where $\dmath^2\bfpar{v}{}{} = \dmath v_R\dmath {\tilde{v}}_{\phi}$. 
The corresponding perturbed potential $\Phi_1(R, \phi, t)$ is 
\begin{align}
\Phi_1(R,\phi,t) \,= &\,\, \Phi_1^{\rm dt}(R, \phi, t) + \Phi_1^{\rm idt}(R, \phi, t)\,,
\label{psn_int}
\end{align}
The term $\Phi_1^{\rm dt}$ is the direct term arising due to gravitational interaction between the disc mass 
particles, and the indirect term $\Phi_1^{\rm idt}$ arises since the coordinate system (centered at the 
massive object) is non-inertial, and is equal to the acceleration of the central mass due to perturbation 
in the disc. These two terms are given by the Poisson integrals
\begin{align}
\Phi_1^{\rm dt}(R, \phi, t) \,=\,& -G\intglwr{0}\intgbth{0}{2\pi} \frac{\siga{1}{}(R',\phi',t) R'\dmath R'\dmath\phi'}
 {\sqrt{R^2+R'^2-2RR'\cos(\phi-\phi')}}\,,\label{phi_dt}
\end{align}
and
\begin{align}
\Phi_1^{\rm idt}(R, \phi, t) \,=\,& GR\intglwr{0}\intgbth{0}{2\pi}\frac{\siga{1}{}(R',\phi',t) \cos(\phi-\phi')
\dmath R'\dmath\phi'}{R'}\,,\nonumber\\
=\,& \pi GR(\delta_{m,1} + \delta_{m,-1})\exp[{\rm i(m\phi - \omega t)}]\int_{0}^{\infty}
 \frac{\siga{a}{}(R')}{R'} \dmath R'\,.
\label{phi_idt}
\end{align}
The second form of the indirect term in the above equation is applicable
if the perturbations are of the form $\siga{1}{}(R',\phi',t) = \siga{a}{}(R')\exp[\rmi(m\phi' - \omega t)]$, 
and is given here for later use. 

The linearised collisionless Boltzman equation (CBE) is given by 
\begin{equation}
\der{f_1}{t} = - \left[f_0, \Phi_1\right]\,,\label{pcbe1}
\end{equation}
where the time derivative on the left hand side is computed along the 
unperturbed orbit, and the bracket $[*,*]$ on the right hand side 
is the Poisson bracket. The solution of linearised CBE is given by 
\begin{equation}
f_1(R,\phi,v_R,\tilde{v}_{\phi},t) = - \int_{-\infty}^{t} \dmath t'\left[f_0, 
\Phi_1\right]_{\bfpar{x'}{}{},\bfpar{v'}{}{},t'}\,,\label{pcbe2}
\end{equation}
where $(\bfpar{x'}{}{},\bfpar{v'}{}{}) = (R',\phi',v_R',\vtilde')$ are 
given by Eqs.~(\ref{epi_approx1}) and (\ref{epi_approx2}). 

We seek solutions of the perturbed quantities for which $\phi$ and $t$ 
dependence of the perturbed quantities goes as $\exp[{\rm i}(m\phi - \omega t)]$. 
The perturbation is assumed to vanish at \mbox{$t \to - \infty$}, which formally 
requires $\omega$ to have a non--zero positive imaginary part however small it may be. 
Expanding the Poisson bracket gives
\begin{align}
\left[f_0,\Phi_1\right] =\, & 
- \DER{f_0}{\bfpar{v}{}{}}\cdot\nabla\Phi_1\,.
\end{align}
\nin
Combining this with Eqn.~(\ref{den_per}), (\ref{psn_int}), and (\ref{pcbe2}) we can 
obtain the perturbed density due to the direct and indirect terms in the potential as
\begin{equation}
\siga{a}{}(R) = \siga{a}{\rm dt}(R) + \siga{a}{\rm idt}(R)\,,\label{pcbe3}
\end{equation}
where
\begin{align}
\siga{a}{*} = \exp[-{\rm i}(m\phi - \omega t)]&\int_{-\infty}^{\infty}\int_{-\infty}^{\infty}
\dmath v_R\dmath{\tilde{v}}_{\phi}\int_{-\infty}^{t} \dmath t' \left[\DER{f_0}{\bfpar{v}{}{}}\cdot\nabla\Phi_1^{*}\right]
_{\bfpar{x'}{}{},\bfpar{v'}{}{},t'}\,, 
\label{pcbe4}
\end{align}
where $*$ stands for `dt' or `idt'.
The next section is dedicated to solving the above set of equations,
by substituting for $\Phi_1^{\rm dt}$ and $\Phi_1^{\rm idt}$ to derive the 
integral--equation for $\siga{a}{}$.

\subsection{The Integral Equation}\label{slf_consistency}

We use the log-spiral expansion of surface density and potential 
\citep{klnjs71,bt08} to write $\Phi_1^{\rm dt}$ 
in Eqn.~(\ref{pcbe4}) in terms of $\siga{a}{}$: 
\begin{equation}
\Phi_1^{\rm \textbf{}dt}(R,\phi,t) = -\frac{G}{R^{1/2}}\int_{-\infty}^{\infty}\frac{\dmath\alpha}{2\pi}
N(\alpha,m)A_m(\alpha)e^{{\rm i}(\alpha q + m\phi - \omega t)}\,,
\label{phi_1dt}
\end{equation}
where $q = \ln R$, and 
\begin{align}
A_m(\alpha) \,&=\,\intgl\dmath q'\,R'^{3/2}\siga{a}{}(R')\rme^{-\rmi\alpha q'}\,,
\label{am}\\
N(\alpha,m)& = \, \pi\frac{\Gamma(z)\Gamma(z^*)}
{\Gamma(z+\frac{1}{2})\Gamma(z^*+\frac{1}{2})}\,,
\label{NM_alpha}
\end{align}
where $z = m/2 + 1/4 + \rmi\alpha/2$. Using this solution for $\Phi_1^{\rm dt}(R, \phi, t)$ and the expression for  
$f_0(R, v_R, \vtilde)$ given in Eq.~(\ref{sch2D}) we get 
\begin{align}
\DER{f_0}{\bfpar{v}{}{}}\cdot\nabla\Phi_1^{\rm dt}{\bigg |}_{\bfpar{x'}{}{},\bfpar{v'}{}{},t'} = 
\frac{G f_0(R, v_R, \vtilde)}{R^{3/2}\sigma_R^{2}}\exp[\rmi(m\phi - \omega t)]&\intgl\frac{\dmath\alpha}{2\pi}\,\times\nonumber\\
\times \left[\left(\rmi\alpha - \frac{1}{2}\right)v_R + \rmi m\gamma^2\vtilde\right]&N(\alpha,m)A_m(\alpha)
\rme^{\rmi\alpha q}{\bigg |}_{\bfpar{x'}{}{},\bfpar{v'}{}{},t'}\,.
\label{pcbe_rhs}
\end{align}
Substituting the above in the expression for $\siga{a}{\rm dt}$ given by 
Eq.~(\ref{pcbe4}) we get
\begin{align}
\siga{a}{\rm dt}(R) = &\frac{G}{R^{3/2}\sigma_R^2}\exp[- {\rm i}(m\phi - \omega t)]\intgl\frac{\dmath\alpha}{2\pi}\,
N(\alpha,m)A_m(\alpha)\times\nonumber\\
&\times \intgl\intgl\dmath v_R \,\dmath\vtilde
\,f_0\intg{t}\dmath t'\,\exp[\rmi(\alpha q' + m\phi' - \omega t')]\times\nonumber\\
&\times \left[v_R'\left({\rm i}\alpha - \frac{1}{2}\right) + {\rm i}m\gamma^2\vtilde'\right]\,.
\end{align}
The above equation has been derived under the standard WKB approximations. 
For details refer to the Appendix K of \citet{bt08}. The main approximations made are:  
\begin{enumerate}
\item We retain terms up to first order in small parameter $|R' - R|$, which is on the order of the epicyclic amplitude. 
\item Also, to a good approximation we can write $|R' - R| \ll R$, and hence  
any slowly varying function of $R'$ such as $\sigma_R(R'),\, \gamma(R'),\, 
\siga{d}{}(R')$, can be replaced by their values at $R$ and taken out of the integral. 
\item Since $q' = \ln(R') = \ln(R + \delta R)$, up to first order $q' = q + (\delta R/R)$.
\item We assume that $|\alpha| \gg m$, and we keep only leading order terms in $\alpha$ 
at each step. This is the equivalent condition to the standard WKB approximation 
as will be proved in the Appendix~\ref{wkbreduction}. For large $\alpha$, the 
leading order radial oscillations of phase are balanced by the (unperturbed) 
drift of $\phi$ at the rate $\Omega$, while the epicyclic drift and 
oscillations of $\phi$ may be neglected. 
Also in the linear term, i.e. the term outside the exponent, only 
$\rmi\alpha v_R'$ term contributes. 
\end{enumerate}

Defining $s = (\omega - m\Omega)/\kappa$, $u = v_R/\sigma_R$ 
and $v = \gamma\vtilde/\sigma_R$, and substituting the expressions for $v_R'$, $\vtilde'$ and $\phi'$ from 
equations~(\ref{epi_approx1})--(\ref{epi_approx2}) in the above integral, we get
\begin{align}
\siga{a}{\rm dt}(R) = &\frac{G\siga{d}{}}{2\pi R^{3/2}\kappa\sigma_R}\intgl\frac{\dmath\alpha}{2\pi} N(\alpha,m)
A_m(\alpha)\,\rme^{\rmi\alpha q}\,\intg{0}\dmath \tau\,\exp\left[-\rmi s\tau \right]{\displaystyle{\mathscr A}}\,,
\label{sigdt1}
\end{align}

\nin
We have defined 
$\tau = \kappa_g(t' - t)$, and
\begin{align}
\mscr{A} = & \intgl\intgl \dmath u\,\dmath v\left(au + bv\right)
\exp\left[-\frac{u^2 + v^2}{2} + \rmi(cu + dv) \right]\,;
\label{A}
\end{align}
\begin{align}
a = &\rmi\alpha\cos\tau\,,\quad
b = \rmi\alpha\sin\tau\,,\quad
c = \frac{\sigma_R}{R\kappa}\alpha\sin\tau\,,\quad
{\text{and}}\quad d = \frac{\sigma_R}{R\kappa}\alpha(1 - \cos\tau)\,.
\label{abcd}
\end{align}
\nin
Since $v_R$ and $\vtilde$ are small on the order of the epicyclic amplitude, 
we replace $\kappa_g$ with $\kappa(R)$ in Eq.~(\ref{sigdt1}).  
Solving the integrals in ${\displaystyle{\mathscr A}}$, it can be brought to the form
\begin{align}
{\displaystyle{\mathscr A}} = - &\frac{2\pi\sigma_R}{R\kappa}\,\alpha^2\sin\tau\,\exp\left[-\chi(1 - \cos(\tau))\right]\,,
\label{a_int2}
\end{align}
\nin
where $\chi =\sigma_R^2\alpha^2/R^2\kappa^2$.
Combining Eqs.~(\ref{sigdt1}) and (\ref{a_int2}), 
expression for $\siga{a}{\rm dt}(R)$ reduces to
\begin{align}
\siga{a}{\rm dt}(R) = & -\frac{G\siga{d}{}}{R^{5/2}\kappa^2}\intgl\frac{\dmath\alpha}{2\pi}\,N(\alpha,m)A_m(\alpha)\,
\alpha^2\,\rme^{\rmi\alpha q}\mscr{I}_1(s,\chi)\,,
\label{sigdt2}
\end{align}
\nin
where
\begin{align}
\mscr{I}_1(s,\chi) = & \intg{0} \dmath\tau\,\sin\tau\exp\left[-\rmi s\tau - \chi\left(1 - \cos\tau\right)\right]\,,
\label{I1I2_1}
\end{align}
\nin
The algebra required to obtain ${\displaystyle{\mathscr I_1}}$ closely 
follows the Appendix-K of \citet{bt08}. The final result is 

\begin{align}
{\displaystyle{\mathscr I_1}}(s, \chi) = &\,-\frac{2\rme^{-\chi}}{\chi}\sum_{n = 1}^{\infty}\left(
\frac{n^2}{n^2 - s^2}\right)I_n(\chi)\,.
\label{I1}
\end{align}

\nin
Substituting for 
${\displaystyle{\mathscr I_1}}$ in Eq.~(\ref{sigdt2}) and simplifying gives
\begin{align}
\siga{a}{\rm dt}(R) =& \,\frac{2G\siga{d}{}}{R^{5/2}\kappa^2}\sum_{n = 1}^{\infty}\left(\frac{n^2}{n^2 - s^2}\right)
\intgl\frac{\dmath\alpha}{2\pi}\,N(\alpha,m)A_m(\alpha)\,\rme^{\rmi\alpha q}\,B_n(\alpha,\chi)\,.
\label{sigdt3}
\end{align}
\nin
Here $B_n(\alpha,\chi)$ is defined as
\begin{align}
B_n(\alpha,\chi) =&\,\, \frac{\alpha^2}{\chi}\,\,\rme^{-\chi}\,I_n(\chi)\,.
\label{BN}
\end{align}
\nin
Note that
\begin{enumerate}
\item $B_n(\alpha, \chi)$ is an even function of $\alpha$. This 
property will be useful while calculating the integral 
over $\alpha$, as we shall see later. 
\item To leading order in $\alpha$, $B_n$ is proportional to $\alpha^2$.
\end{enumerate}

Having expressed $\siga{a}{\rm dt}$ in the desired form given by Eqn.~(\ref{sigdt3}), 
we now turn to the calculation of $\siga{a}{\rm idt}$\,. Below we prove 
that to leading order $\siga{a}{\rm idt} = 0$.  
$\Phi_1^{\rm idt}(R,\phi,t)$, as given in 
Eq.~(\ref{phi_idt}), can be rewritten as
\begin{align}
\Phi_1^{\rm idt}(R,\phi,t) = &\, R\exp[\rmi(m\phi - \omega t)]\mscr{J}_m\,,\nonumber\\
{\text{where,}}\,\,\,\,\,\,\,\,\mscr{J}_m = \pi G(\delta_{m,1}& + \delta_{m,-1})
\intglwr{0}\frac{\dmath R'}{R'}\siga{a}{}(R')\,, 
\end{align}
\nin
is a constant. Using this we can write
\begin{align}
\DER{f_0}{\bfpar{v}{}{}}\cdot\nabla\Phi_1^{\rm idt}{\bigg |}_{\bfpar{x'}{}{},\bfpar{v'}{}{},t'} = 
-\frac{f_0}{\sigma_R^{2}}\exp[\rmi(m\phi' - \omega t')]\left(v_R' + \rmi m\gamma^2\vtilde'\right)\mscr{J}_m\,.
\label{sigidt1}
\end{align}
\nin
Combining Eqs.~(\ref{pcbe4}) and (\ref{sigidt1}) and defining 
$s$, $\tau$, $u$ and $v$ as done before $\siga{a}{\rm idt}$ becomes
\begin{align}
\siga{a}{\rm idt} = & -\frac{\mscr{J}_m\siga{d}{}}{2\pi\kappa\sigma_R}\intgbth{-\infty}{0}\dmath \tau 
\rme^{-\rmi s\tau}\mscr{A'}\,,
\label{sigidt2}
\end{align}
where
\begin{align}
\mscr{A'} = & \intgl\intgl \dmath u\,\dmath v\left(a'u + b'v\right)
\exp\left[-\frac{u^2 + v^2}{2}\right]\,, 
\end{align}
\begin{align}
a' = & \cos\tau - \rmi m\gamma\sin\tau\,,\nonumber\\
b' = & \sin\tau + \rmi m\gamma\cos\tau\,.
\label{abcd_1}
\end{align}
\nin
In writing the above integral we have neglected the oscillations in 
$\phi$ 
and the epicyclic drift term. Neglecting these terms involves the 
same level of approximation as in calculating $\siga{a}{\rm dt}$. 
The integral in $\mscr{A'}$  is exactly equal to zero since the integrand 
is an odd functions of $u$ and $v$, therefore 
\begin{equation}
\siga{a}{\rm idt} = 0\,.
\label{sigidt4}
\end{equation}
As given in Eq.~(\ref{pcbe3}), $\siga{a}{}$ is the sum of both 
$\siga{a}{\rm dt}$ and $\siga{a}{\rm idt}$. Combining Eqs.~(\ref{pcbe3}), 
(\ref{am}), (\ref{sigdt3}), and (\ref{sigidt4}) we get 
\begin{align}
\siga{a}{}(R) = &\, \frac{G\siga{d}{}}{R^{5/2}\kappa^2}\intgl\dmath q'\,
\mathcal{G}_m(s,\chi,q - q') R'^{3/2}\siga{a}{}(R')\,,
\label{int_eqn1}
\end{align}
where
\begin{align}
\mathcal{G}_m(s,\chi,q) = &\, 2\sum_{n = 1}^{\infty}\left(\frac{n^2}{n^2 - s^2}\right)
\intgl\frac{\dmath\alpha}{2\pi}\,N(\alpha,m)\,B_n(\alpha,\chi)\,\rme^{\rmi\alpha q}\,.
\label{gm1}
\end{align}

We have already discussed that both the functions $N(\alpha,m)$ and 
$B_n(\alpha,\chi)$ 
are even functions of $\alpha$. Since `$\sin(\alpha q)$' and `$\cos(\alpha q)$' 
are odd and even functions of `$\alpha$' respectively, only `$\cos$' term in the 
integral over $\alpha$ survives. Combining these, the integral equation 
reduces to,

\begin{align}
\mathcal{S}(R)= &\,\intgl\dmath q'\,\left[\frac{\mathcal{C}(R)\mathcal{C}(R')}{\kappa(R)}
\mathcal{G}_m(s,\chi,q - q')\right]\mathcal{S}(R')\,,
\label{int_eqn2}
\end{align}
where
\begin{align}
 \mathcal{G}_m(s,\chi,q) = &\, 4\sum_{n = 1}^{\infty}\left(\frac{n^2}{n^2 - s^2}\right)
\intglwr{0}\frac{\dmath\alpha}{2\pi}\,N(\alpha,m)\,B_n(\alpha,\chi)\,\cos(\alpha q)\,.
\label{gm2}
\end{align}
and we have defined 
\begin{align}
\mathcal{S}(R) =&\, \frac{R^{3/2}\siga{a}{}(R)}{\mathcal{C}(R)} \qquad{\text{and}}\qquad
\mathcal{C}(R) = \sqrt{\frac{G\sigd{}(R)}{R\kappa(R)}}\,.
\label{defS(R)}
\end{align}
Equation~(\ref{int_eqn2}), together with the functions defined 
in Eq.~\eqref{gm2} \& \eqref{defS(R)} is the integral eigenvalue 
problem for tightly-wound linear modes of an axisymmetric disc in 
the epicyclic approximation. The application of this equation is not 
restricted to Keplerian discs,  it could also be used to explore modes 
of non-Keplerian discs such as galactic disc. We show in Appendix A 
that the local limit of this equation gives the same dispersion relation 
as is used for stellar discs \citep{bt08}.

\section{The slow mode limit}\label{slwmode}
The general integral equation, Eqn.~\ref{int_eqn2}, derived in the previous section
is difficult to solve in its standard form due to the presence of an infinite series. 
The near equality of $\Omega$ and $\kappa$ for nearly Keplerian discs has a 
simplifying effect on this equation. For such discs, 
$\Omega(R) = \kappa(R) + \pom$, and $\pom \sim  \mathnormal{O}(\varepsilon)$. 
If we make an ansatz that the eigenfrequency $\omega \sim  \mathnormal{O}(\varepsilon) \ll 1$, 
then we find that  $s \simeq -m$, $\gamma 
\simeq 2$ to leading order and 
\begin{equation}
 m^2 - s^2 = \frac{2m(\omega - m\pom)}{\kappa}\,.
\end{equation}
Therefore, in the summation over $n$ from $1$ to $\infty$ in 
Eq.~(\ref{gm2}), $n = m$ term 
dominates due to the presence of the factor $1/(n^2 - s^2)$. Note that 
there is no restriction on $m$, and slow modes exist for all $m$. The 
magnitude of the frequencies obtained later in the paper indeed 
satisfies this ansatz, thereby validating the slow mode approximation.

Further simplification is possible if we take $\sigma_R(R) = \sigma R\kappa(R)$, 
where $\sigma$ is a dimensionless constant less than unity. Using such a profile 
means that the radial profile of $\sigma_R$ is same as 
$R\kappa(R) \simeq R\Omega(R) = v_c(R)$. The epicyclic condition is satisfied 
for $\sigma < 1$, which is one of the fundamental assumption that has gone 
into deriving the integral equation. Moreover, \citet{jt12} have studied a 
specific problem with this simplifying assumption for $\sigma_R$. Therefore, 
it is convenient to use this in our formulation to compare results obtained 
by them using our integral equation for validation of our equation.

Since we have assumed $\sigma$ to be constant, $\chi$ is also 
a constant, leading to considerable simplification. Using these 
simplifying assumptions in Eqs.~(\ref{int_eqn2})--(\ref{defS(R)})  we obtain
\begin{align}
\omega\,\mathcal{S}(R) = &\, m\pom(R)\mathcal{S}(R) + \intgl\dmath q'\,\mathcal{H}_m(\sigma,q,q')\,\mathcal{S}(R')\,,
\label{eig1}
\end{align}
\nin
where the kernel 
\begin{align}
 \mathcal{H}_m(\sigma,q,q') = &\, 2m\,\mathcal{C}(R)\mathcal{C}(R')
\intglwr{0}\frac{\dmath\alpha}{2\pi}\,N(\alpha,m)\,B_m(\alpha,\chi)\,\cos(\alpha(q - q'))\,.
\label{hm}
\end{align}
\nin
This is the integral eigenvalue problem for slow modes in a nearly 
Keplerian collisionless disc. Note that the kernel of the 
integral equation is symmetric in $R$ and $R'$ (or $q$ and $q'$). Hence 
the integral operator on the RHS can be regarded as a linear Hermitian 
operator. Properties of Hermitian operators imply that the eigenvalues 
$\omega$ are all real, {\it thus all slow modes are stable}. 
Also, the eigenfunctions $\mathcal{S}(R)$ can be assumed to be real. 

\section{Numerical method}\label{num_mthd}

In this section we discuss the numerical methods used to solve the above derived 
integral equation. First step is to solve for $\mathcal{H}_m(\sigma, q, q')$. 
We define the function $\mathit{K}_m$    
\begin{equation}
\mathit{K}_m(\sigma,q) = \intglwr{0}\frac{\dmath\alpha}{2\pi}\,N(\alpha,m)\,B_m(\alpha,\chi)\,\cos(\alpha q)\,,
\label{km}
\end{equation}
in terms of which the kernel $\mathcal{H}_m(\sigma,q,q')$ takes the form
\begin{equation}
\mathcal{H}_m(\sigma,q,q') = 2m\,\mathcal{C}(R)\mathcal{C}(R') \mathit{K}_m(\sigma, q - q')\,.
\end{equation}
\nin
Calculation of $\mathit{K}_m$ involves integral over $\alpha$, the function 
$N(\alpha,m)$, and  $B_m(\alpha,\chi)$.

The functional form of $B_m(\alpha,\chi)$ 
contains the combination ${\rm e}^{-\chi}I_m(\chi)$ and we calculate this instead of 
$I_m$ because $I_m$ increases exponentially for large value of its argument.
Next we tabulate $N(\alpha, m)$ as a function of $\alpha$ for a given value of $m$. 
Using the identities $\Gamma(z^*) = \Gamma(z)^*$ and $zz^* = |z|^2$, the expression 
for $N(\alpha,m)$ in Eq.~(\ref{NM_alpha}) becomes
\begin{align}
N(\alpha,m) = &\,\pi\left|\frac{\Gamma(z)}{\Gamma(z+\frac{1}{2})}\right|^2\,.
\end{align}
We then use the following identity to calculate $|\Gamma(x + \rmi y)|$:
\begin{align}
|\Gamma(x + \rmi y)| = &\, |\Gamma(x)|\prod_{n = 0}^{\infty}\left[\frac{(x + n)^2 + y^2}{(x + n)^2}\right]^{-1/2}\,.
\end{align}
Taking $\log$ on both sides converts the product into summation. 
Convergence of series over $n$ is achieved iteratively till the 
accuracy of $10^{-8}$ is attained, and then we exponentiate to obtain 
$\Gamma(x + \rmi y)$. Also $|\Gamma(x + \rmi y)|$ is a decreasing 
function of $y$, and since we need to evaluate $|\Gamma(x + \rmi y)/\Gamma(x + \rmi y + 1/2)|$ 
as a function of $y$, when the minimum of machine 
precision is reached, $N(\alpha,m)$ becomes indeterminate. 
We avoid this by using the asymptotic form of this function given in 
Eq.~(\ref{NM_aysmtotic}).

The function $\mathit{K}_m$ is next calculated by evaluating the integral 
over $\alpha$ by using the Gaussian quadrature. Although the integrand becomes 
small for large values of $\alpha$, but due to the presence of oscillatory 
cosine function the tail of the integrand has to be handled carefully. The integral 
evaluated up to infinity ensures complete cancellation due to oscillatory 
functions, however since we evaluate the integral numerically up to a finite  
range in $\alpha$, we have to deal with small spurious contributions from the 
tail. To avoid this we take $\alpha$ range to be quite high. For typical 
maximum value of $\alpha_{\rm max} = \rme^{25-27}$, and number of grid points, 
in $\log(\alpha)$ scale, $N_{\alpha} = 10^7$, the function $\mathit{K}_m$ 
converges to sixth or seventh decimal place. 


\begin{figure}
\begin{center}
\includegraphics[angle=-90,scale=0.64]{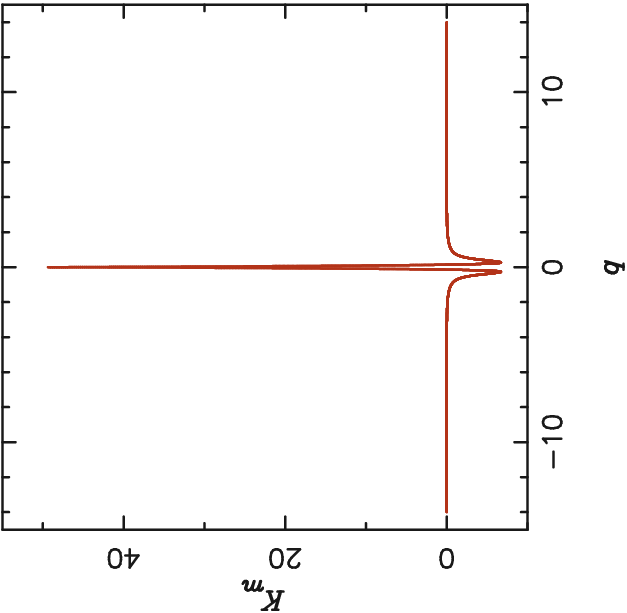}
\includegraphics[angle=-90,scale=0.64]{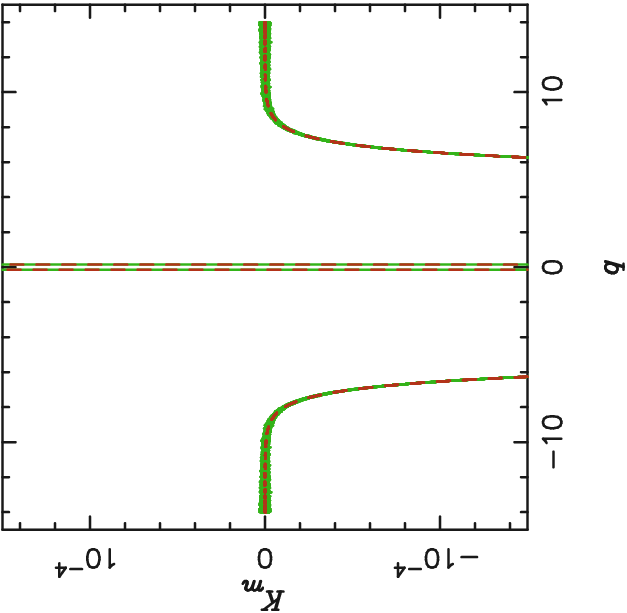}
\caption{Plot of $\mathit{K}_m$ vs $q$ for $m = 1$, $\sigma = 
\sigma_R/R\kappa(R) = 0.1$. Left panel is the plot of 
smoothened $\mathit{K}_m$ and the right panel displays the zoom 
of $y$-axis for a window of $-1.5\times10^{-4}$ to 
$1.5\times10^{-4}$. Green one is the original curve 
whereas red one is after doing a box-smoothening.}
\label{hmvsq}
\end{center}
\end{figure}


In our calculation the variables $q$ and $q'$ are assigned the range 
$[-6\,, 6]$, which is divided into $4000$ points each. The lower range 
of $q$ is chosen to avoid numerical singularities for $q \to - \infty$. 
Also, the upper limit is chosen since the surface density decreases 
substantially beyond $q = 6$. These enter
$\mathit{K}_m$ in the combination $q - q'$ which then ranges from $-12$ 
to $12$. We first tabulate $\mathit{K}_m$ as a function of `$q$' ranging 
from $-14$ to $14$ with a grid size of $10^5$ for a given value of 
`$m \,\,\& \,\,\sigma$' as defined in Eqn.~(\ref{km}). 
The range is extended from $[-12, 12]$ to $[-14, 14]$ 
just to make sure that tail effects are minimized. As we have 
discussed earlier there is numerical noise in the tail of the 
integral over $\alpha$ due to the presence of oscillatory functions. 
To reduce the noise we do a box smoothening for $\mathit{K}_m$, which 
works quite well. In Fig.~\ref{hmvsq} we display a plot of $\mathit{K}_m$ 
as a function of $q$. The left panel is the plot of box smoothed 
$\mathit{K}_m$ and the right panel is zoom of $y$-axis from 
$-1.5\times10^{-4}$ to $1.5\times10^{-4}$\,. The green curve is the 
unsmoothed curve and the red curve is the smoothed curve. The reduction 
in numerical noise can be easily seen in this figure.

Having once tabulated $\mathit{K}_m$ as a function of $q$ for 
a given $m$ and $\sigma$, we divide $-6 \le q$ (and $q'$) $\le 6$ 
into a grid of $n_q$ points and interpolate the tabulated function 
to calculate actual matrix entries. Further, the calculation of 
$\mathcal{H}_m$, once we have $\mathit{K}_m$, involves calculation of 
simple algebraic functions only. The discretization of the 
integral over $q'$ in Eqn.~(\ref{eig1}) follows the scheme 

\begin{equation}
\int_{-\infty}^{\infty} {\mathrm{d}q'}\,
\mathcal{H}_m(\sigma,q_i,q')\,\mathcal{S}(q') \quad\longrightarrow\quad \sum_{j=1}^{n_q}w_{q_j}\,\mathcal{H}_m(\sigma,q_i,q_j)\,
\mathcal{S}(q_j)\,,
\label{disct_schm}
\end{equation}
\nin
where we have divided $q$ (and $q'$) on a grid of $n_q$ points using the 
Gaussian quadrature rule, and $w_{q_j}$ are the appropriate weights. Using 
this, the discretized integral-equation can be written as
\begin{align}
{\bf A}\mathcal{S} = &\, \omega\mathcal{S},
\end{align}
where ${\bf A}$ is a $n_q \times n_q$ matrix defined as
\begin{align}
{\bf A} = &\, \Big[w_{q_j}\mathcal{H}_m(\sigma,
q_i, q_j) + m\pom_i\delta_{ij}\Big].
\end{align}
Row and column indices are $i$ and $j$, respectively. Note that 
no summation is implied over repeated indices.  
The presence of unequal weights makes the matrix non-symmetric. 
Since the weights are all positive, the symmetry is easily restored by the
transformation given in $\S~18.1$ of \citet{prs92}. We write $\tilde{\mathcal{H}}_m = \mathcal{H}_m\,{\bf D}$, where ${\bf D} 
= {\rm diag}(w_{q_j})$. Now 
\begin{align}
 {\bf D}^{1/2}\,\tilde{\mathcal{H}}_m\,{\mathcal{S}} = & \left({\bf D}^{1/2}\,\mathcal{H}_m\,{\bf D}^{1/2}\right)\,
{\bf D}^{1/2}\,{\mathcal{S}},\nonumber\\
= & \left({\bf D}^{1/2}\,\mathcal{H}_m\,{\bf D}^{1/2}\right)\,{\bf h},
\label{krnl:rstr}
\end{align}
\nin where ${\bf h} = {\bf D}^{1/2}\,{\mathcal{S}}$ and ${\bf D}^{1/2} = 
{\rm diag}(\sqrt{w_{q_j}})$.  We use this 
as our input to calculate the eigenvalues and eigenfunctions (which now is 
${\bf h}$) numerically rather than $\tilde{\mathcal{H}_m}$, which is originally 
there in matrix ${\bf A}$. And then restore ${\mathcal{S}}$  by using the 
transformation $\mathcal{S} = {\bf D}^{-1/2}\,{\bf h}$, where 
${\bf D}^{-1/2} = {\rm diag}(1/\sqrt{w_j})$. We have used the linear algebra 
package LAPACK \citep{lapack} to calculate eigenvalues and eigenvectors.  

\section{Numerical results}\label{num_sol}

\begin{figure}
\begin{center}
\includegraphics[angle=-90,scale=0.64]{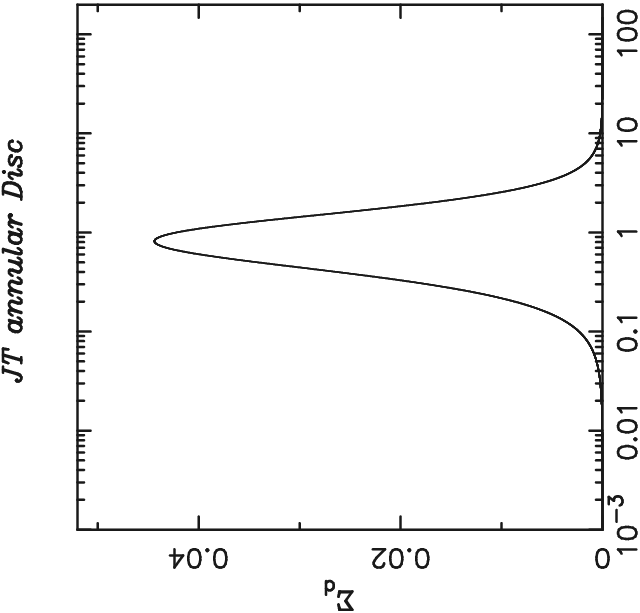}
\includegraphics[angle=-90,scale=0.64]{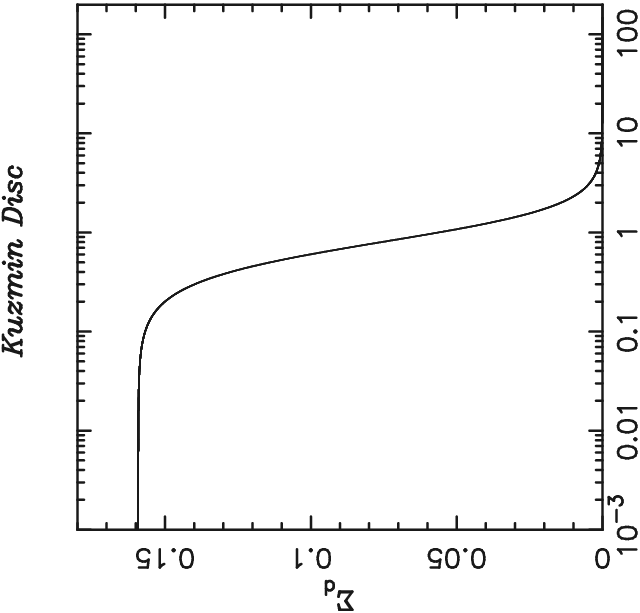}
\includegraphics[angle=-90,scale=0.64]{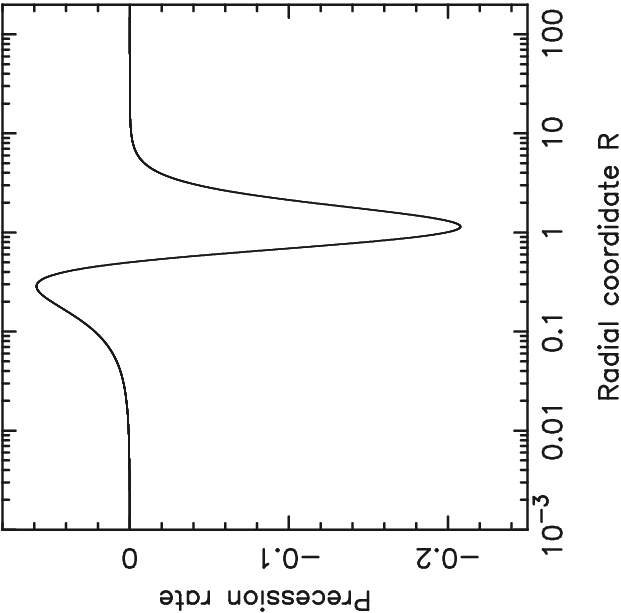}
\includegraphics[angle=-90,scale=0.64]{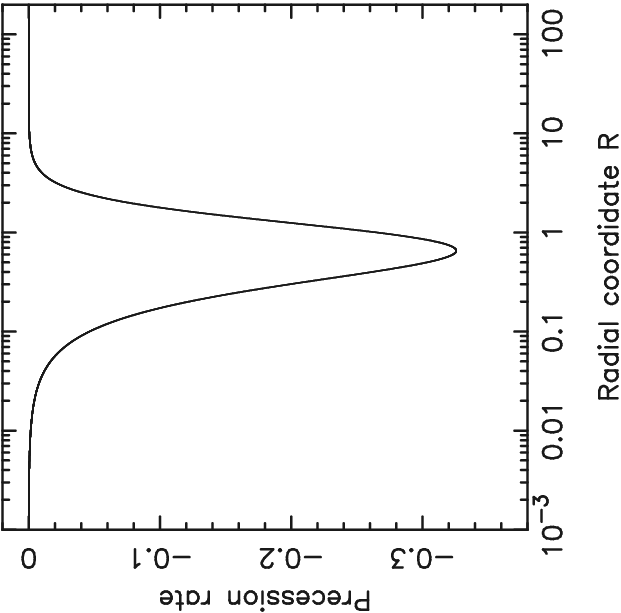}
\caption{Surface density and 
precession frequency profiles for JT annular disc and Kuzmin 
disc are displayed. Panels on the left correspond to JT annular 
disc and those on the right are the profiles for Kuzmin disc. Top 
row displays the surface density, $\sigd{}$, and the bottom  
one are the plots for the precession rate $\pom$.}
\label{kuzvstmre}
\end{center}
\end{figure}

We consider two contrasting models of the disc density to explore the possible 
eigenvalues and eigenfunctions of the slow modes. Both the models contain a 
characteristic disc scale-length `$a$', which we use to cast the equations 
in a dimensionless form. $R/a$ is the dimensionless radius; and to convert 
other physical quantities to dimensionless form we use $\md/a^2$ as the 
characteristic surface density and $\Omega^* = \sqrt{GM/a^3}$ as the characteristic 
orbital frequency. The net effect is that the dimensionful eigenfrequencies $\omega$ 
are obtained from the dimensionless frequencies by multiplying with $(\Omega^*a^3/G\md)^{-1}$. 
In the rest of the paper the notation ( $R$, $\sigd{}$, $\omega$, $\Omega$ and $\mathcal{C}(R)$, 
etc), used earlier for dimensionful quantities, will stand for dimensionless quantities. The 
two discs models considered for our numerical exploration are:
\begin{enumerate}
\item {\bf JT annular disc}: This is an annular disc model around the central 
massive object obtained by subtracting two 
Toomre discs \citep{tmre63}. This profile was analyzed for slow modes by 
\citet{jt12} by solving the collisonless Boltzmann equation in the ring-ring 
interaction approximation. Since the eigenvalues for this problem are known, 
this model also serves to validate our eigenequation. Following them we call 
the disc JT annular disc. The radial profile (dimensionless form) is given by 
\begin{equation}
\sigd{\rm JT}(R) = \frac{3}{4\pi}\left[\frac{1}{(1 + R^2)^{3/2}} - \frac{1}{(1 + R^2)^{5/2}}\right] 
= \frac{3R^2}{4\pi(1 + R^2)^{5/2}}\,,
\label{sig_tmre}
\end{equation}
And the corresponding precession rate for nearly circular orbits is 
\begin{equation}
\pom^{\rm JT}(R) = \frac{3(1 - 4R^2)}{4\Omega(R)(1 + R^2)^{7/2}}\,.
\label{pom_tmre}  
\end{equation}
Note that $\Omega(R)$ used here is the dimensionless azimuthal frequency. 
$\pom^{\rm JT}(R) > 0$ for $0 < R < 1/2$, zero at $R = 1/2$ and negative 
thereafter. Positive and negative maxima are $0.05861$ and $-0.2078$, 
respectively. Both $\sigd{\rm JT}(R)$ and $\pom^{\rm JT}(R)$ are plotted in 
the left pannel of Fig.~\ref{kuzvstmre}.
 \item {\bf Kuzmin disc:} Several earlier investigations of 
 slow modes \citep{tre01,st10,gss12} have considered the Kuzmin disc model. 
 We consider this model to make comparison with the earlier works. The 
 surface density and the precession frequency for a Kuzmin disc are 
\begin{align}
\sigd{\rm Kz}(R) \,=\, & \frac{1}{2\pi(1 + R^2)^{3/2}}\,,\\
\pom^{\rm Kz}(R) \,=\, & \frac{-3}{2\Omega(R)(1 + R^2)^{5/2}}\,.
\end{align}
Both the quantities are in dimensionless units. Note that 
$\pom^{\rm Kz} \le 0$ for all values of $R$. We plot both 
$\sigd{\rm Kz}(R)$ and $\pom^{\rm Kz}(R)$ in the right panel of 
Fig.~\ref{kuzvstmre}.
\end{enumerate}

\nin

These profiles differ from each other: (1) Surface density for Kuzmin disc 
is centrally concentrated whereas for JT annular disc is concentrated 
about $R = 1$. (2) Precession frequency is negative throughout for Kuzmin 
disc, whereas for JT annular disc it starts from zero, attains a positive 
maxima, becomes negative, reaches a minimum, and then goes to 
zero. We give and compare the results from both these profiles next. 


\begin{figure}
\begin{center}
\includegraphics[angle=0,scale=1.2]{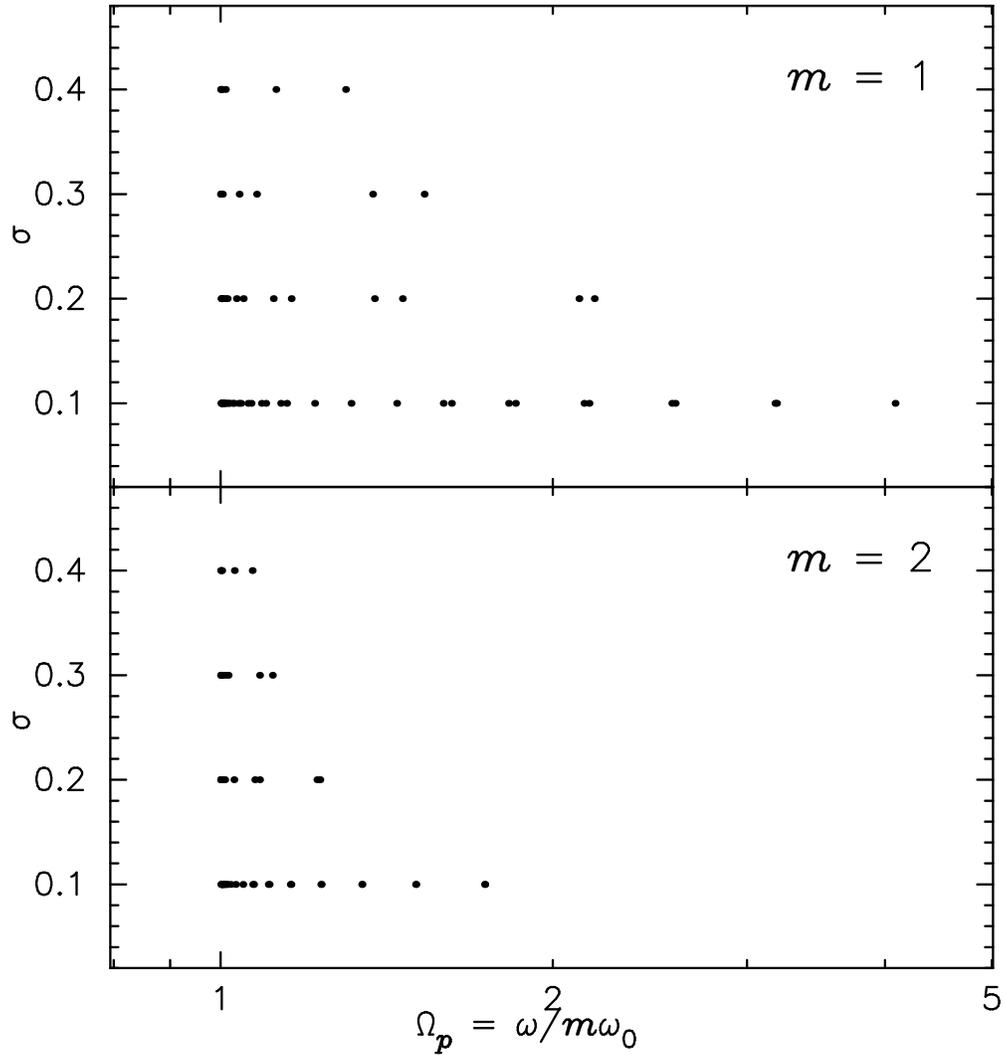}
\caption{Non-singular eigenfrequencies for JT annular disc profile for 
$m = 1$ \& $2$ and $\sigma = 0.1, 0.2, 0.3$ and $0.4$. Eigenvalues 
are all real and prograde. X-axis is $\Omega_p = \omega/m\omega_0$, where 
$\omega_0$ is the positive maxima of precession rate and Y-axis 
is the $\sigma$ value. Plots are labelled for their respective $m$ 
values.}
\label{eig_plot_tmre}
\end{center}
\end{figure}


\subsection{JT annular Disc}\label{toomre}
We present results for  $m = 1$ and $2$,  with $\sigma = 0.1$, $0.2$, $0.3$ \& $0.4$. 
Both continuous as well as discrete eigenfrequencies are supported by the eigenequation. 
The continuous spectrum, however, has singular eigenfunctions where the eigenfrequencies 
are $\omega=m\pom$. The plot of the more interesting discrete eigenvalues for JT annular 
disc is given in Fig.~\ref{eig_plot_tmre}. The X-axis is the dimensionless pattern 
speed, $\Omega_p = \omega/m\omega_0$, where $\omega_0 = 0.05861$ is positive maxima 
of precession frequency. Note that all 
frequencies are measured in the units of `$\varepsilon\sqrt{GM/a^3}$\,'\,, the 
natural slow mode frequency. The Y-axis is $\sigma$, which is a 
dimensionless measure of the heat in the disc. We note the following 
trends from Fig.~\ref{eig_plot_tmre} for $m = 1$ and $m = 2$ modes: 
\begin{enumerate}
\item The modes are all stable with prograde pattern speeds 
$\Omega_p > 1$.
\item For given $(\sigma, m)$, the pattern speed belongs to a 
discrete spectrum. Let $\Omega_{\rm max}(\sigma, m)$ be the largest 
eigenvalue of this spectrum. Then
\begin{enumerate}
\item At fixed $m$, $\Omega_{\rm max}$ is a decreasing function 
of $\sigma$.
\item At fixed $\sigma$, $\Omega_{\rm max}(\sigma,1) > \Omega_{\rm max}(\sigma,2)$.
\end{enumerate}
\end{enumerate}

Plot for $m = 1$ is to be compared with Fig.~$4$ of \citet{jt12}. 
Mean eccentricity used by the authors is linearly proportional 
to $\sigma$ used in the present work. Apart from the last property 
of the eigenspectra mentioned above (about which nothing has been 
said by the authors), our conclusions are consistent with their 
results. Eigenvalues match within a few percent which Jalali \& Tremaine 
get by solving collisionless Boltzmann equation, and even better with 
the eigenvalues obtained after solving the local WKB dispersion relation. 
All the eigenvalues with $\Omega_p \le 
1$ that we get are singular for all the values of $\sigma$.


\begin{figure}
\begin{center}
\includegraphics[angle=0,scale=0.58]{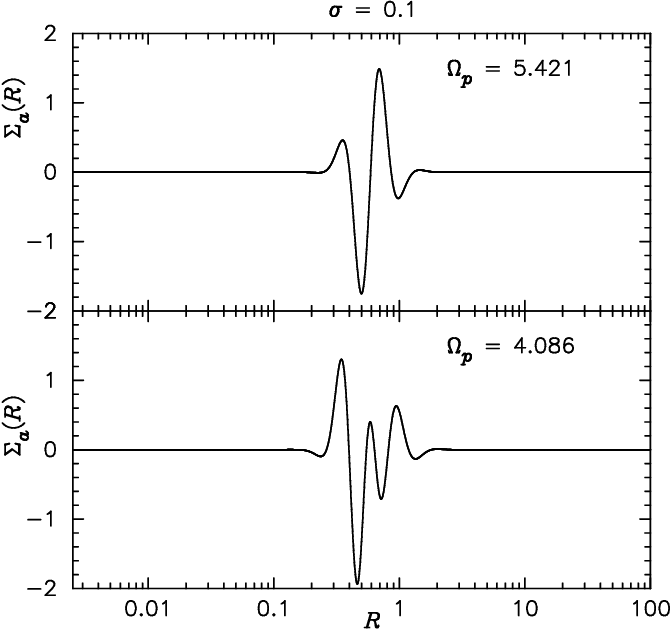}
\includegraphics[angle=0,scale=0.58]{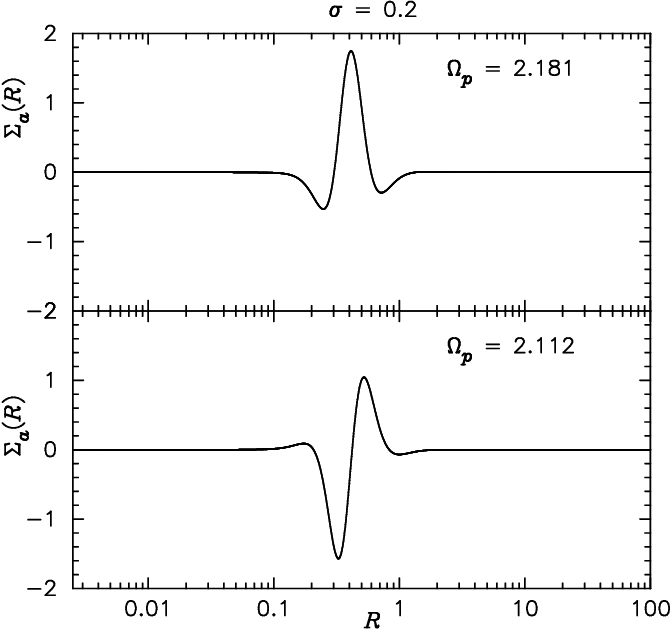}
\includegraphics[angle=0,scale=0.58]{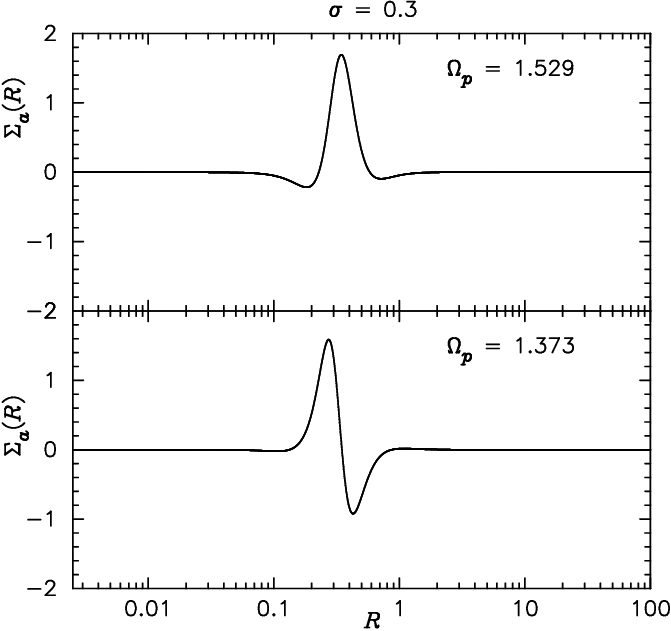}
\includegraphics[angle=0,scale=0.58]{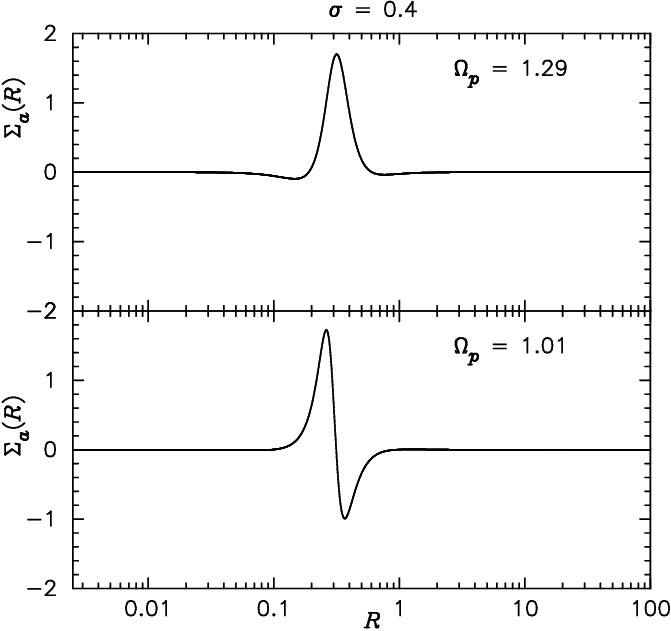}
\caption{Plot of perturbed surface density, $\siga{a}{}(R)$ 
as a function of $R$ for $m = 1$ and JT annular disc profile. 
Eigenfunctions for first two eigenvalues for each value of $\sigma$ 
are displayed. Panels are labelled for the values of $\sigma$ 
and $\Omega_p$. Functions are square normalized to unity such 
that $\int\dmath q\,\siga{a}{2}(R) = 1$.}
\label{eig_plot_tmre_m=1}
\end{center}
\end{figure}


\begin{figure}
\begin{center}
\includegraphics[angle=-90,scale=0.58]{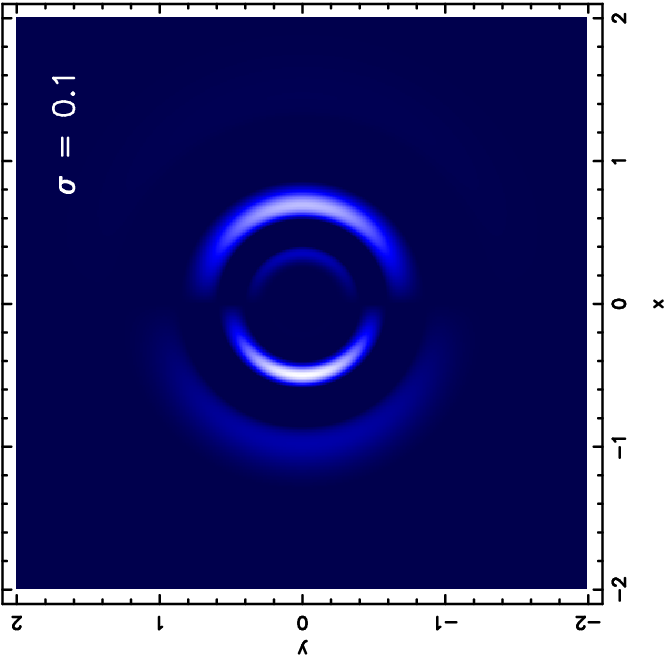}
\includegraphics[angle=-90,scale=0.58]{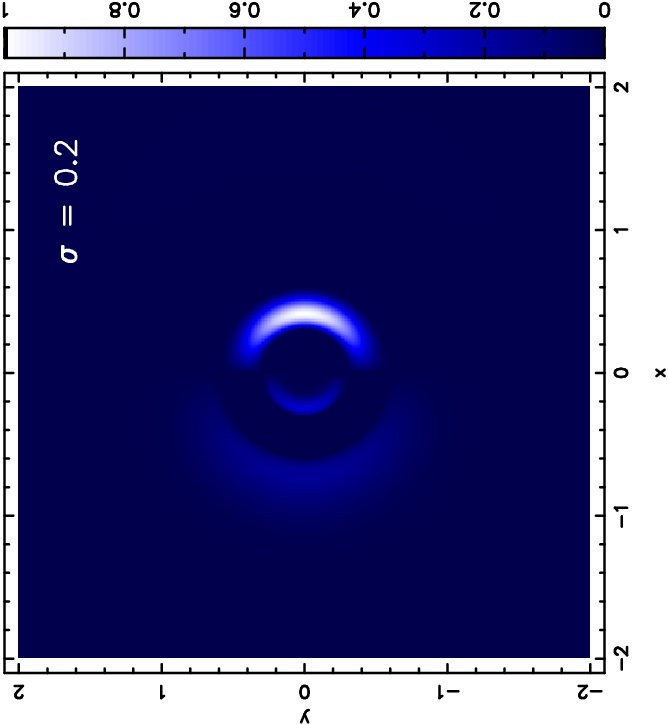}
\includegraphics[angle=-90,scale=0.58]{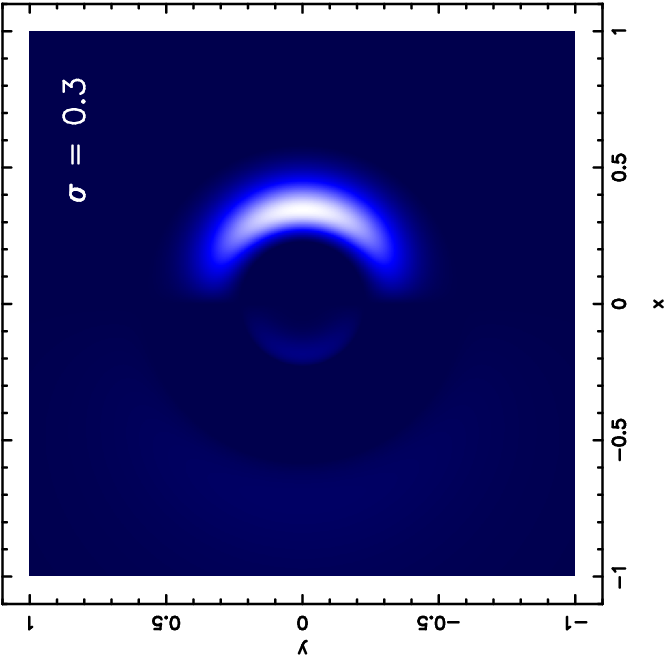}
\includegraphics[angle=-90,scale=0.58]{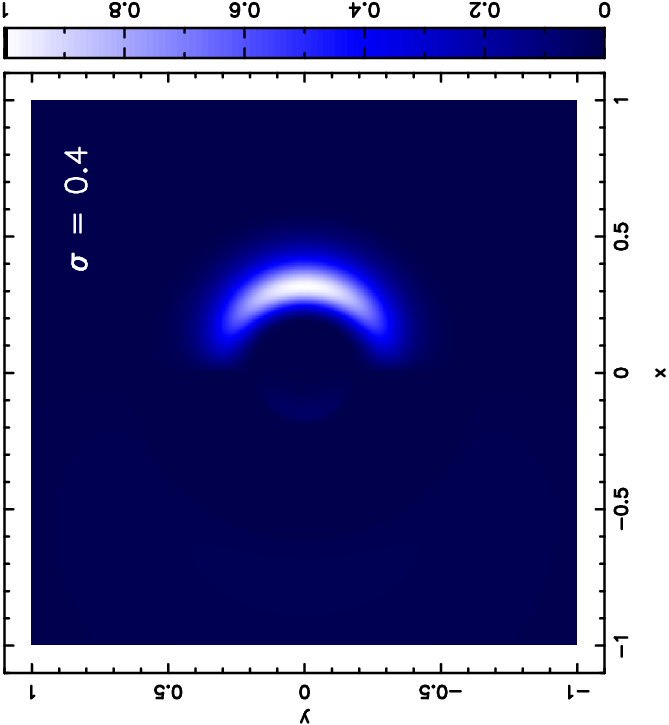}
\caption{Patterns of oscillatory waves for JT annular disc. We 
have displayed the positive component of Real part of 
$\siga{1}{}(R,\phi,t)$ at $t = 0$, for $m = 1$. Plots are for 
highest value of $\Omega_p$ for each value of $\sigma$ and the 
panels are labelled for its respective $\sigma$ values. 
The surface density is square normalized to unity in all panels 
and the color scheme for contours from $0$ to $1$ is plotted 
in a wedge on right side of the image.}
\label{sigplot_tmre}
\end{center}
\end{figure}


\begin{figure}
\begin{center}
\includegraphics[angle=0,scale=0.58]{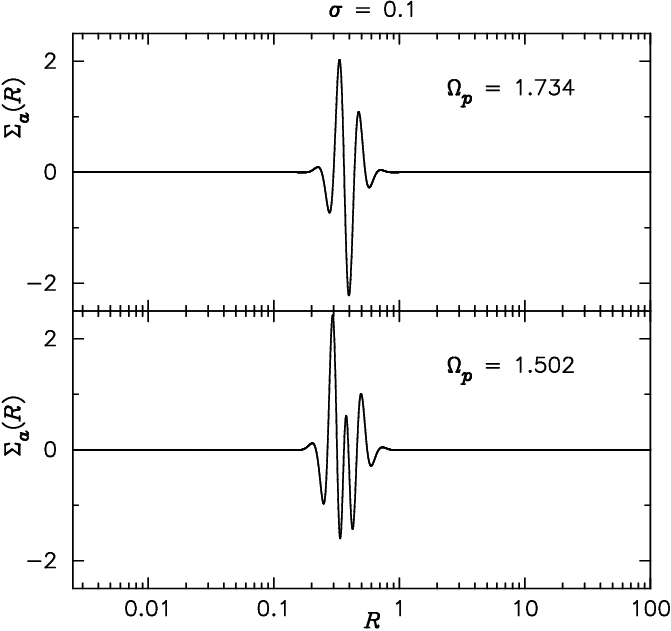}
\includegraphics[angle=0,scale=0.58]{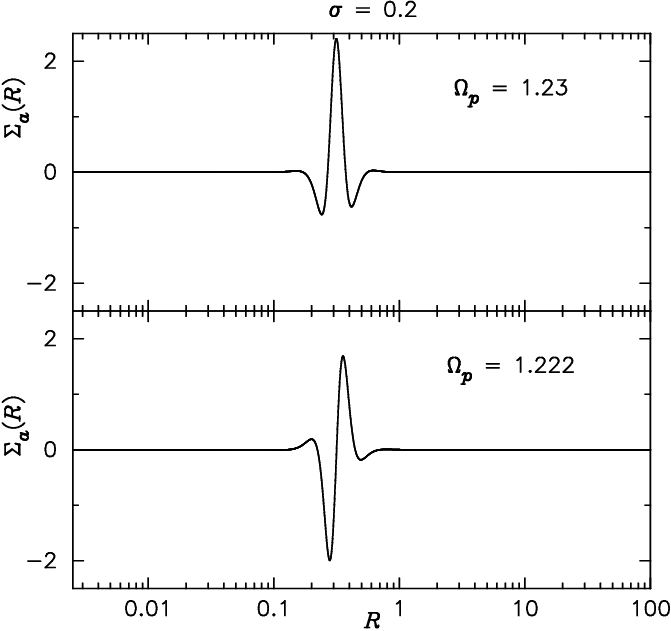}
\includegraphics[angle=0,scale=0.58]{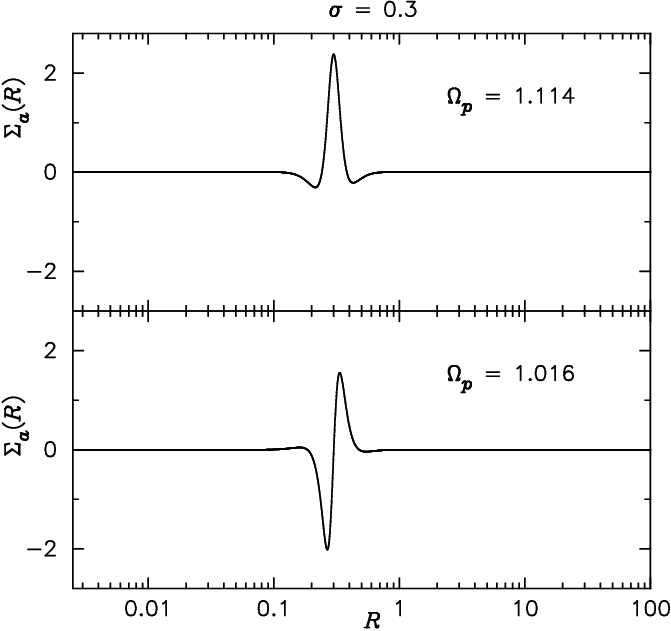}
\includegraphics[angle=0,scale=0.58]{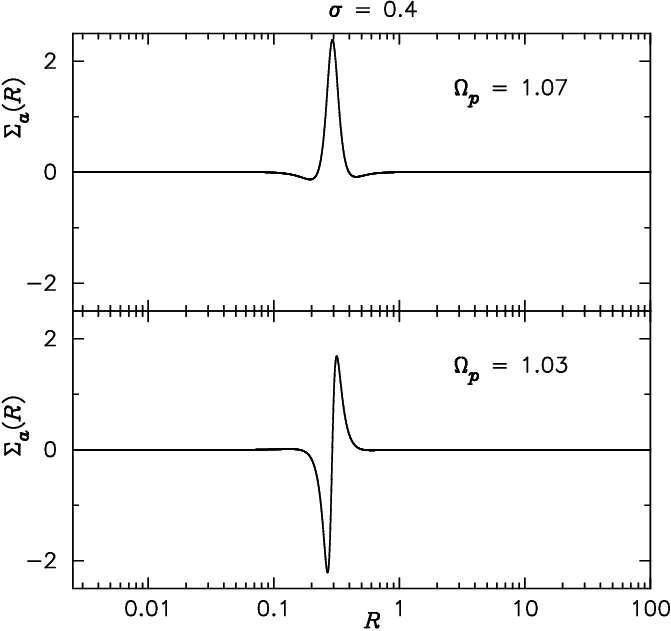}
\caption{$\siga{a}{}(R)$ vs $R$ plot for $m = 2$ for two eigenmodes
with least number of nodes, using $\sigd{\rm JT}$ as the unperturbed 
disc. Plots are labelled for their respective $\sigma$ and 
$\Omega_p$ values.}
\label{eig_plot_tmre_m=2}
\end{center}
\end{figure}


\begin{figure}
\begin{center}
\includegraphics[angle=-90,scale=0.58]{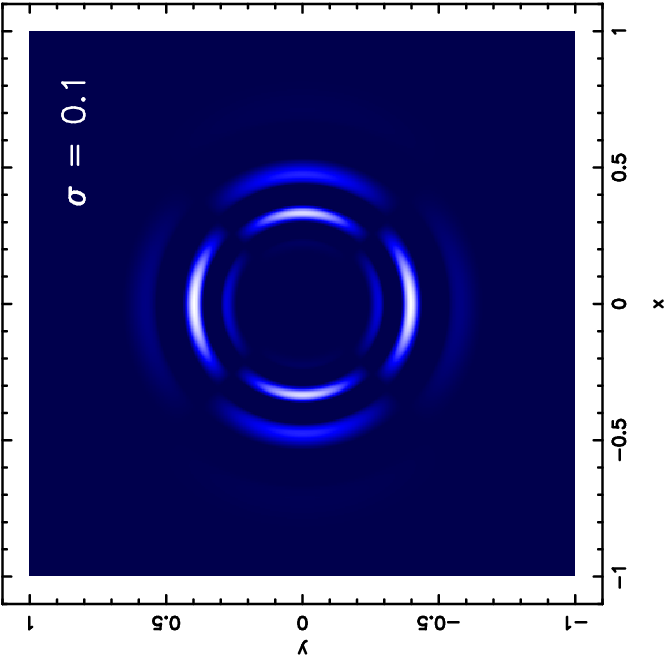}
\includegraphics[angle=-90,scale=0.58]{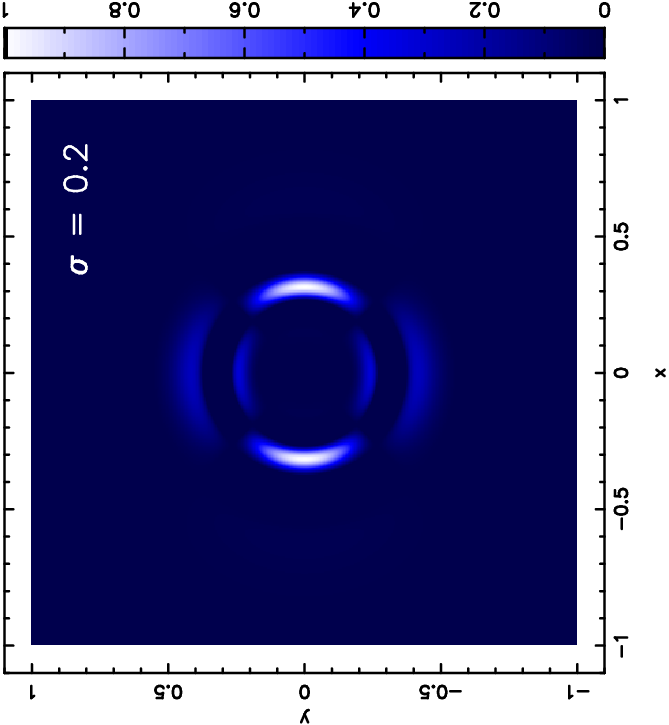}
\includegraphics[angle=-90,scale=0.58]{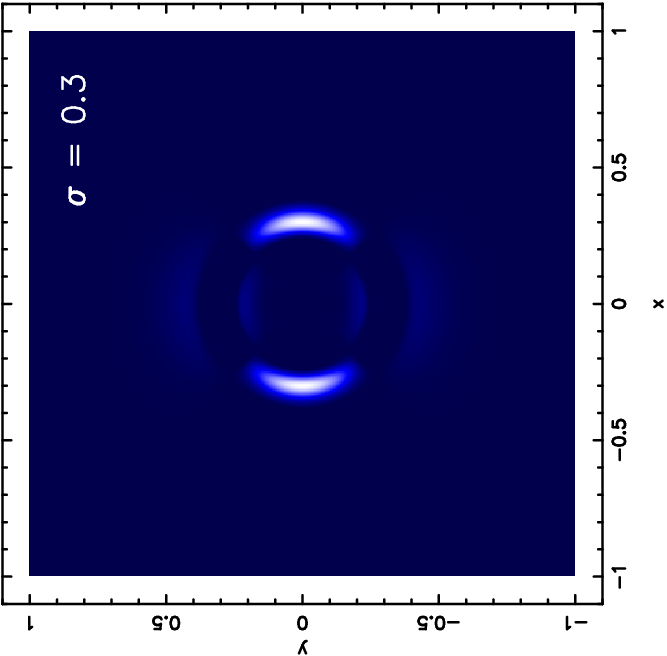}
\includegraphics[angle=-90,scale=0.58]{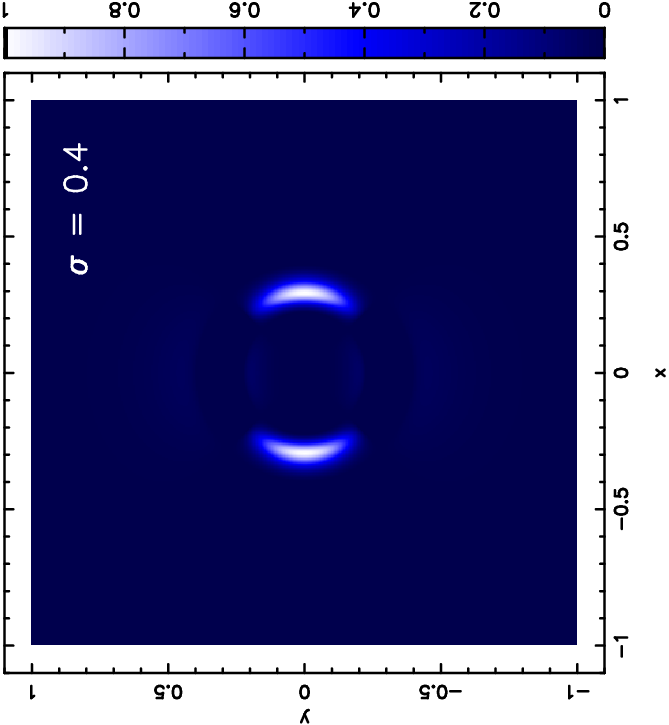}
\caption{Image of positive component of Real part of 
$\siga{1}{}(R,\phi,t)$ at $t = 0$ for the same set of parameters 
as in figure~\ref{eig_plot_tmre_m=2}. Plots of the largest 
$\Omega_p$ for each $\sigma$ value are displayed.}
\label{sigplot_tmre_m=2}
\end{center}
\end{figure}


Figure~\ref{eig_plot_tmre_m=1} and \ref{eig_plot_tmre_m=2} 
show the radial profile of $\siga{a}{}(R)$ for $m = 1$ and $2$, respectively. 
Functions are normalized such that $\int\dmath q\,\siga{a}{2}(R) = 1$. 
We plot the eigenfunctions for the first two eigenvalues for all the 
values of $\sigma$. Panels are labelled for the values 
$\sigma$ and $\Omega_p$. Number of nodes increase 
as the $\Omega_p$ value decreases. In Fig.~\ref{sigplot_tmre} 
and \ref{sigplot_tmre_m=2} we plot the image of oscillatory 
patterns of the 
positive part of $\siga{a}{}(R)\cos(m\phi)$ for $m = 1$ and $2$, 
respectively, which is essentially the positive component 
of the real part of $\siga{1}{}(R,\phi,t)$ at $t = 0$.   
Plots for the highest values of $\Omega_p$ are displayed and their 
respective $\sigma$ values are given in the panels. Surface density 
is normalized to unity in all the panels. Contours range from 
$0$ to $1$, and the corresponding colors are shown in a wedge on the 
right side. Wavepackets are more radially compact for lower values 
of $\sigma$. 

\subsection{Kuzmin disc}\label{kuzmin}
\begin{figure}
\begin{center}
\includegraphics[angle=0,scale=1.2]{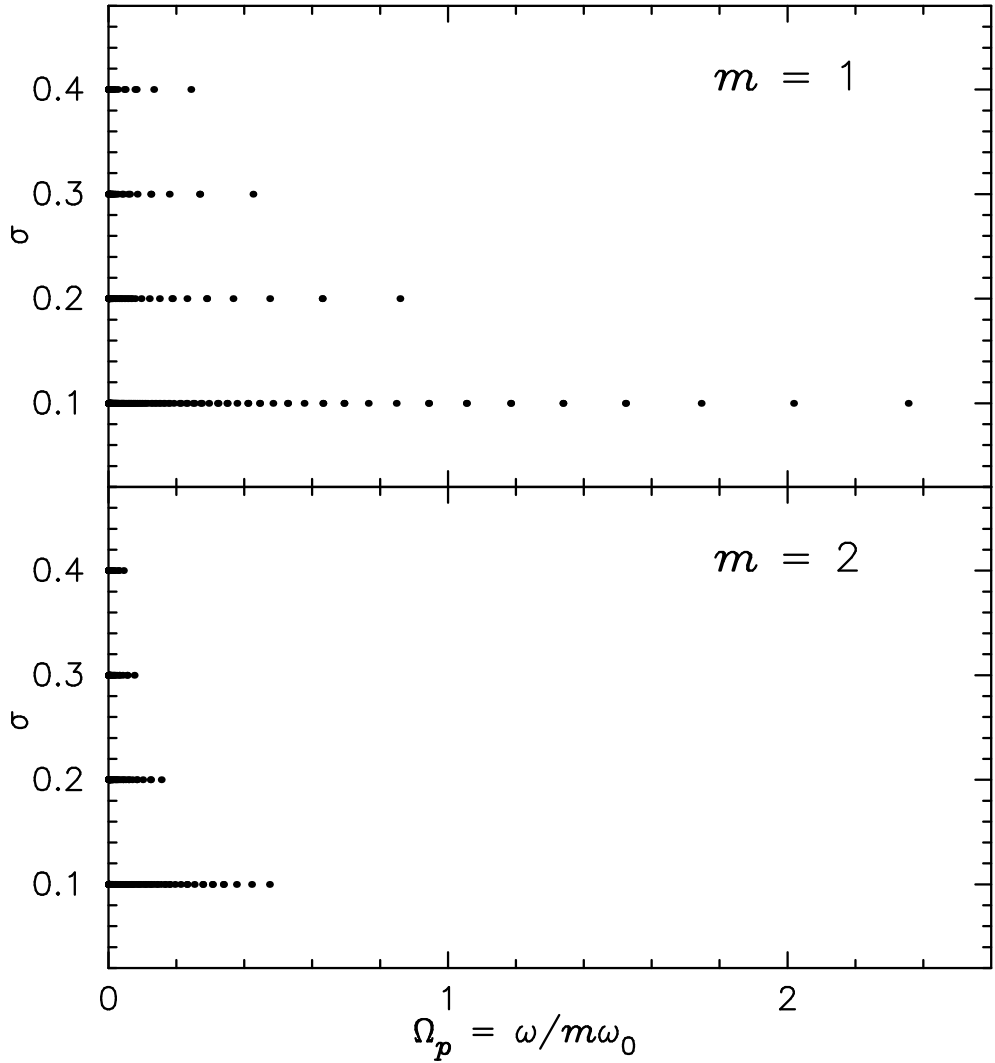}
\caption{Plot of eigenvalues for Kuzmin disc profile. Horizontal 
axis is the pattern speed $\Omega_p = \omega/m\omega_0$, where 
$\omega_0$ is the maximum of $|\pom|$ and the vertical axis is the 
$\sigma$ value. Panels are labelled for their respective $m$ 
values.}
\label{eigplot_kzmn}
\end{center}
\end{figure}

In this subsection we present the results for Kuzmin disc profile. Eigenspectrum 
we get in this case also is composed of singular modes given by $\omega = m\pom$ 
and the non-singular eigenvalues. 
Figure~\ref{eigplot_kzmn} gives the plot of non-singular 
eigenvalues for the Kuzmin disc. We have plotted for $m = 1$ \& $2$ 
and $\sigma = 0.1, 0.2, 0.3$ and $0.4$. Horizontal axis is 
$\Omega_p = \omega/m\omega_0$, where $\omega_0$ is the maxima of $|\pom|$ rather than 
positive maxima (as used for the JT disc) because for Kuzmin disc $\pom(R) \le 0$. 
We note the following trends in the eigenspectrum for $m = 1$ and 
$m = 2$:
\begin{enumerate}
\item Eigenmodes are stable with prograde pattern speeds $\Omega_p > 0$, 
in contrast to JT disc where $\Omega_p > 1$.
\item For a given value of $m$, $\Omega_{\rm max}$ is a decreasing 
function of $\sigma$ and $\Omega_{\rm max}(\sigma, 1) > 
\Omega_{\rm max}(\sigma, 2)$. Variation of $\Omega_{\rm max}$ with 
$\sigma$ and $m$ is similar to the JT disc.
\end{enumerate}
 
\begin{figure}
\begin{center}
\includegraphics[angle=-90,scale=1.2]{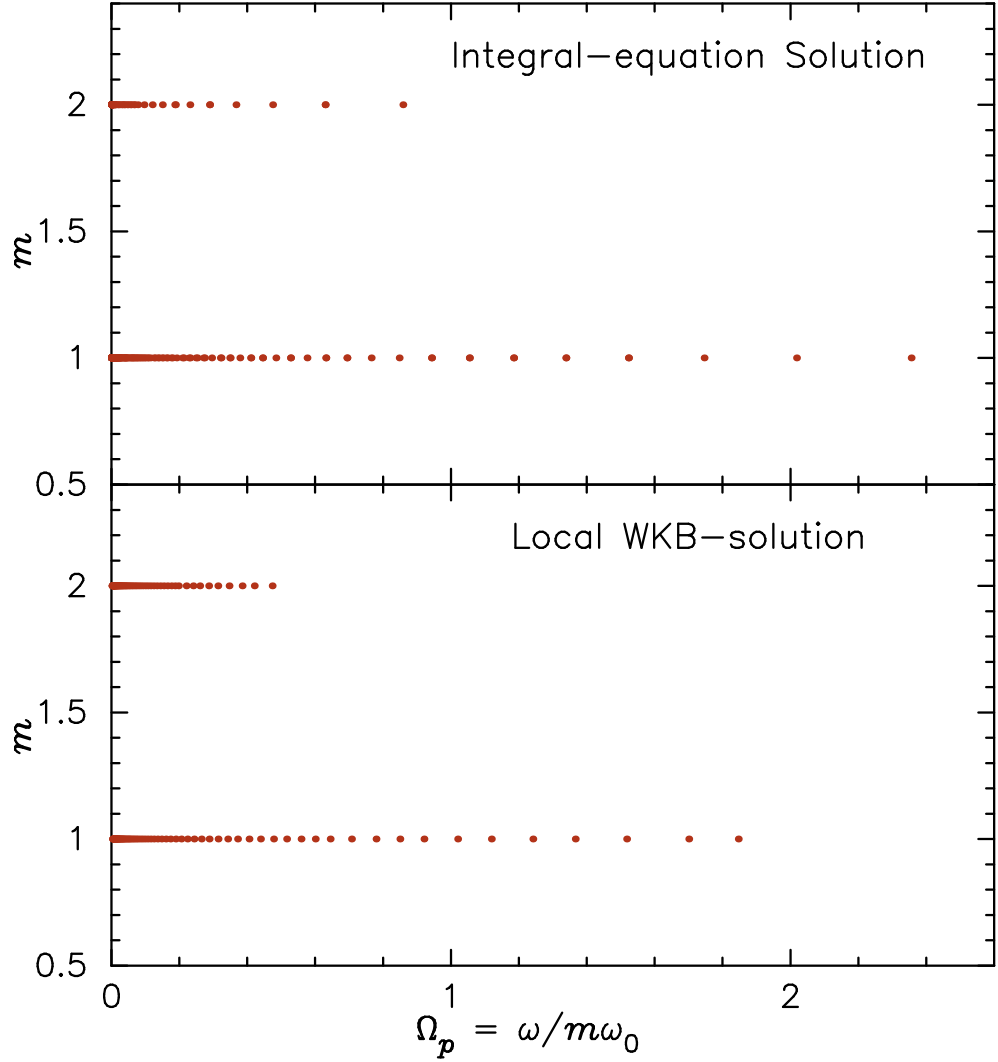}
\caption{Comparison between the eigenvalues obtained by solving the 
local WKB dispersion relation and the 
eigenvalues calculated using the integral-equation for Kuzmin 
disc, with $\sigma = 0.1$. Horizontal and vertical axis are 
eigenvalue, $\Omega_p$ and $m$, respectively.}
\label{kzmn_wkb_comp}
\end{center}
\end{figure}

We also solve the local WKB-dispersion relation (as given in the Appendix 
of \citet{jt12}) for the Kuzmin disc model. 
In Fig.~\ref{kzmn_wkb_comp} we compare the solution of local 
WKB dispersion relation 
and the eigenmodes calculated in this section for $\sigma = 0.1$ 
Top panel gives the integral equation solution and the lower panel 
gives eigenvalues obtained from local WKB dispersion relation. 
The eigenvalues differ from each other 
by about $20$\%, but qualitative trends are the same; for example, 
as we increase the value of $m$, the $\Omega_p$ value decreases, 
and $\Omega_p$ values increases with decreasing $\sigma$. Second 
one can be seen by comparing the plots for other $\sigma$ values.

\begin{figure}
\begin{center}
\includegraphics[angle=0,scale=0.58]{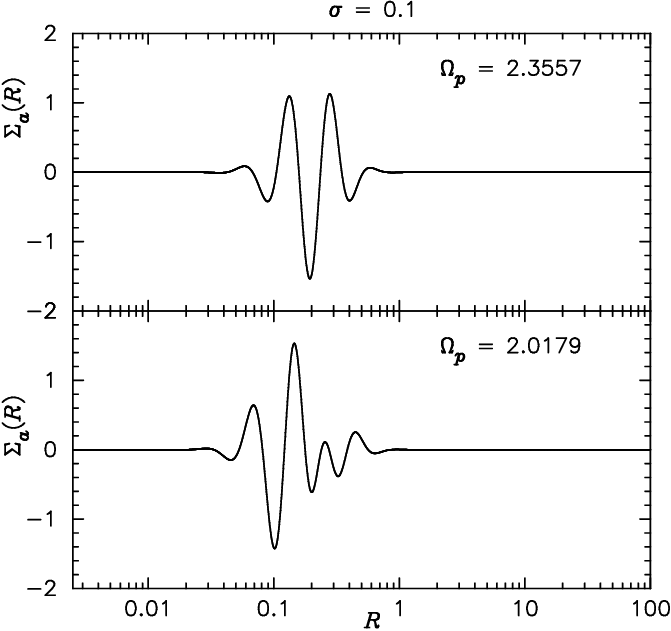}
\includegraphics[angle=0,scale=0.58]{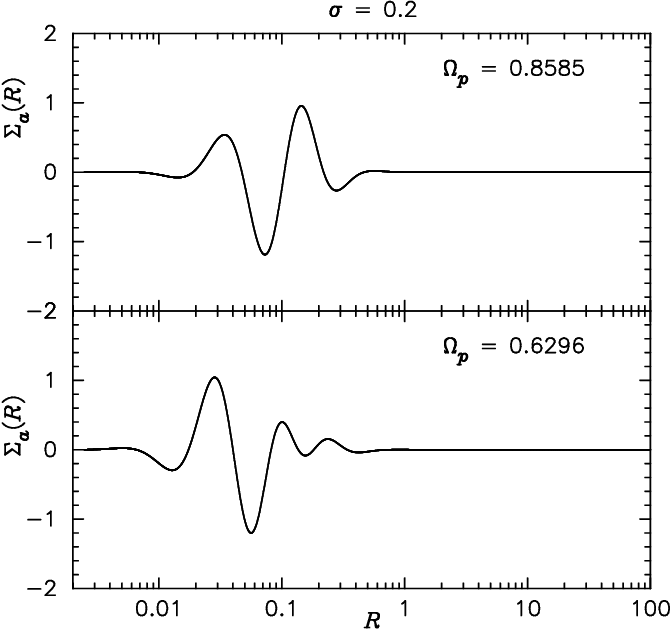}
\caption{$\siga{a}{}(R)$ vs $R$ plot for $m = 1$, with 
$\sigd{\rm Kz}$ as the unperturbed density. Panels are labelled 
for their respective $\sigma$ and $\Omega_p$ values. Normalization 
for $\siga{a}{}(R)$ is same as used for JT annular disc.}
\label{eig_plot_kzmn_m=1}
\end{center}
\end{figure}

\begin{figure}
\begin{center}
\includegraphics[angle=-90,scale=0.58]{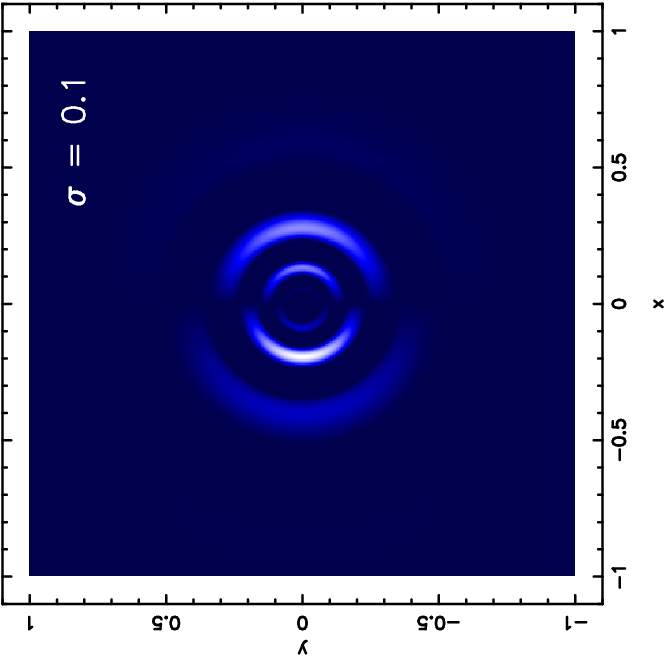}
\includegraphics[angle=-90,scale=0.58]{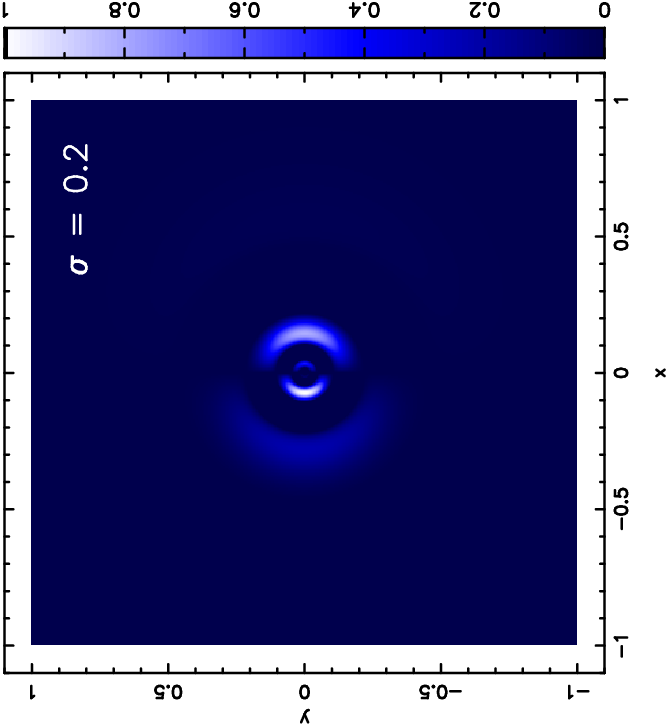}
\caption
{Image of positive part of the real component of 
$\siga{1}{}(R,\phi,t)$ at $t = 0$, for highest eigenvalue 
for each $\sigma$ value for the plots displayed in 
Fig.~\ref{eig_plot_kzmn_m=1}.}
\label{sigplot_kzmn}
\end{center}
\end{figure}

\begin{figure}
\begin{center}
\includegraphics[angle=0,scale=0.58]{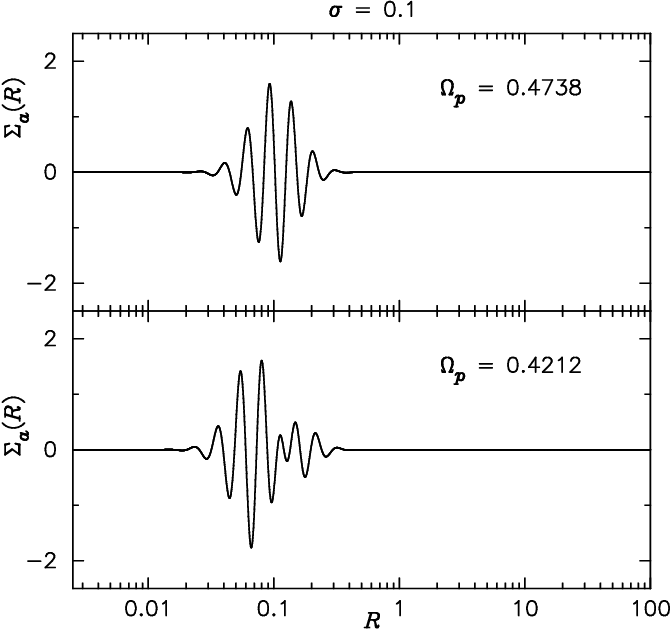}
\includegraphics[angle=0,scale=0.58]{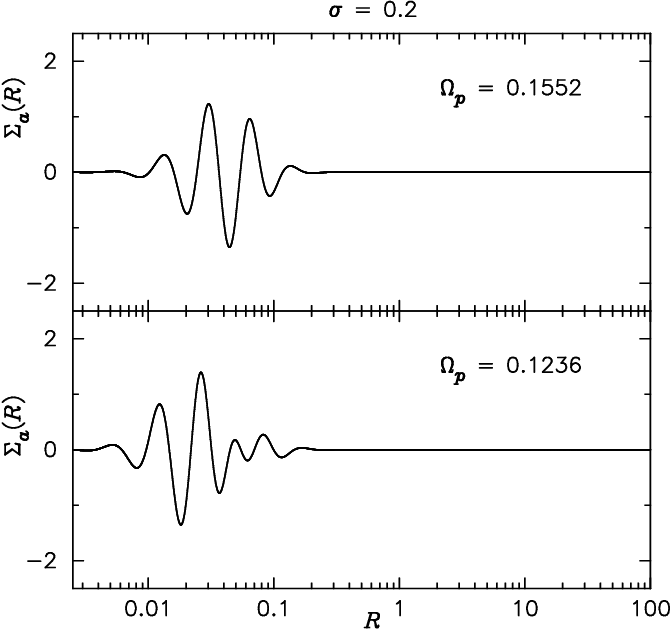}
\caption{Square normalized eigenfunction $\siga{a}{}(R)$ 
vs $R$ plot for $m = 2$ for Kuzmin disc profile. Relevant 
labelling for $\sigma$ and $\omega$ value is given in the plot.}
\label{eig_plot_kzmn_m=2}
\end{center}
\end{figure}

Figure~\ref{eig_plot_kzmn_m=1}, \ref{sigplot_kzmn} and 
\ref{eig_plot_kzmn_m=2} are the plots of the perturbed surface 
density of the Kuzmin disc profile. In Fig.~\ref{eig_plot_kzmn_m=1} 
we the plot $\siga{a}{}(R)$ as a function $R$ for $m = 1$. Plots are 
labelled for their $\sigma$ and $\Omega_p$ values. $\siga{a}{}(R)$ is 
normalized such that $\int \dmath q\,\siga{a}{2}(R) = 1$. 
Figure~\ref{sigplot_kzmn} give images of density 
enhancement region, real part of $\siga{1}{}(R,\phi,t)$, at $t = 0$ 
for the highest eigenvalue for each $\sigma$ value used in 
Fig.~\ref{eig_plot_kzmn_m=1}. Color scheme and normalization used 
is same as that used in Fig.~\ref{sigplot_tmre}.  
Radial profile of square normalized 
wave functions $\siga{a}{}(R)$ for $m = 2$ are given in 
Fig.~\ref{eig_plot_kzmn_m=2}. Apart from the exact forms 
of $\siga{a}{}$ for all cases, overall properties of the eigenfunctions 
are same as that we get using JT annular disc; (1) Number of nodes 
increase with decreasing $\Omega_p$ value, (2) wavefunctions are 
radially more compact for lower values of $\sigma$. 

Next we compare the eigenvalues we get by solving the 
integral equation for the Kuzmin disc in this section with the 
solution of local WKB dispersion relation and the 
integral-equation solution for softened-gravity disc studied 
in \citet{tre01}. The model for 
velocity dispersion used in the present study directly corresponds 
to the one used by \citet{tre01}. In Table~\ref{tab:comp} we 
give the first five non-degenerate eigenvalues for all the three 
studies for $\sigma = 0.1$ and $m = 1$ (softening length 
$\beta = 0.1$ as used by \citet{tre01}). First column is the 
local WKB-solution, second and third columns are for integral equation 
solution for softened-gravity disc and the collisionless discs, 
respectively. The eigenvalues match within $\sim 10 \%$ and the match 
is within few percent for higher number of nodes, where the WKB 
approximation is expected to work better. 


\begin{table}
\centering
\begin{center}
\begin{tabular}{|c|c|c|} 
\hline
Solution for&Integral equation&Integral equation\\ 
local WKB dispersion&solution for&solution for\\
relation&softened gravity disc&collisionless disc\\
\hline
0.601&0.67&0.767\\
0.554&0.62&0.657\\
0.494&0.57&0.569\\
0.445&0.52&0.496\\
0.404&0.48&0.436\\
\hline
\end{tabular}
\caption{Table of comparison between the solution of local WKB
dispersion relation for 
Kuzmin disc, Integral equation solution 
for softened-gravity disc \citep{tre01} and Integral equation 
solution studied in the present chapter for Kuzmin disc. 
Values are for $\sigma = 0.1$.}
\label{tab:comp}
\end{center}
\end{table}


\section{Conclusions}\label{cnclsn}

We have formulated  linear perturbations in an axisymmetric collisionless stellar disc as 
an eigenvalue problem. By linearising the collisionless Boltzmann equation, we have derived 
an eigenvalue equation in the tight winding limit. We go a step further than the canonical 
WKB dispersion relation by treating the density-potential relation non-locally. This 
formalism allows us to determine both the eigenfrequencies as well as eigenfunctions for a 
stellar disc. We expect the accuracy of eigenvalues obtained through this formalism to be 
comparable to the WKB eigenvalues, which are fairly reasonable estimates of the eigenvalues 
as shown by \citet{jt12}, but the advantages are: (1) We are able to obtain the eigenfunctions 
to a good accuracy, (2) and our formalism is considerably simpler than that of \citet{jt12}. 

Although our formulation is applicable to all stellar discs, for this work we have used it 
only to analyse the slow modes of a nearly Keplerian disc. We have calculated numerically the 
slow modes for two different unperturbed surface density profiles, namely: (1) JT annular disc, 
(2) and the Kuzmin disc. Radial profile of velocity dispersion was assumed to be 
$\sigma_R(R) = \sigma R\kappa(R)$, where $\sigma < 1$ is a constant. This is a reasonable 
model for velocity dispersion \citep{jt12}. Our conclusions for the slow modes of these two discs are:

\begin{itemize}
\item Since the kernel of the slow mode integral-eigenvalue problem is symmetric, therefore all the 
eigenvalues are real. Moreover all the non-singular eigenvalues are prograde, $\Omega_p > 0$. 

\item The important trends seen by varying  $\sigma$ are: (a) Largest eigenfrequency is a 
decreasing function of $\sigma$, (b) and the number of non-singular eigenvalues increases as 
$\sigma$ decreases. 

\item $\Omega_{\rm max}(\sigma, m)$ value decreases as we go from $m = 1$ to $2$. In addition, 
for a given $\Omega_p$ value, number of nodes for $m = 1$ are larger than 
that for $m = 2$. In other words eigenfunctions are more radially 
compact for $m = 1$.

\item The general behaviour of the eigenfunctions is that: (1) The wavelength of 
oscillations decreases with decreasing pattern speed, (2) the number of nodes increase 
with decreasing $\Omega_p$ values, (3) and wavefunctions are radially more compact 
for lower values of $\sigma$. 

\item  Largely the behaviour of the eigenfrequencies and the eigenfunctions 
is similar for the two unperturbed surface density chosen, but there 
are quantitative differences, such as the values of pattern speeds. In the case 
of Kuzmin disc all $\Omega_p > 0$ are found to be non-singular, although there is a 
continuum of eigenvalues close to $\Omega_p = 0$; whereas in the case of 
JT annular discs the eigenvalues with $\Omega_p \le 1$ are all singular modes.
\end{itemize}
\nin

These conclusions are consistent with the earlier works of \citet{tre01, gss12, jt12}. 
Since the slow modes are stable, the excitation mechanism for such modes 
is important; for example, a close encounter with a passing, massive object can act as 
an external perturbation that can excite these modes. \citet{jt12} have 
considered such a phenomenon in detail and conclude that external perturbation
is an excellent mechanism to excite the slow modes.    

Slow modes exists with arbitrary azimuthal wavenumber $m$ but 
the modes with lower $m$ values are large scale and hence are 
most prominent in the observations. Also lower $m$ modes are 
easy to excite, for example by an external perturber 
\citep{jt12}. As noted by \citet{jt12}, galactic discs surrounding 
a suppermassive BH and debris disc around stars are similar in the 
sense that dynamics of both the discs are influenced by the central object 
(star/suppermassive BH) and the self-gravity of the disc. Hence 
the analysis presented in this paper is also applicable to debris 
disc. \citet{jt12} proposed that most of the non-axisymmetric 
features in the debris disc may be 
due to slow modes. There are other hypothesis like a presence of 
massive planet in debris discs, which can also cause these 
asymmetries in the discs. These can be distinguished from slow 
modes if the structures are observed for long enough time or with 
higher resolutions. Features due to slow modes will rotate much 
slower as compared to the angular speed of the disc whereas structures 
due to planets in the discs will rotate at a speed comparable to 
the angular speed. 

Double peak 
stellar distribution is observed in two galaxies: M$31$ and 
NGC$4486$B. Distribution in both these galaxies differ from each 
other, for instance, both the peaks in NGC$4486$B are symmetric 
w.r.t. the photocenter in contrast to the peaks in M$31$. Double 
peak stellar distribution in NGC$4486$B is more likely to be due 
to $m = 2$ modes rather than $m = 1$ eccentric modes for M$31$. 
Both these galaxies being different 
morphologically can excite different $m$-modes predominantly. These 
eccentric modes may also play an important role in feeding the 
central BH in galaxies.

\appendix
\section{The local limit}\label{wkbreduction}
Here we verify that a local approximation--valid when $|\alpha|$ is 
not just much larger than $m$, but is truly large--to the integral 
problem reduces it to the well-known WKB dispersion relation of 
\citet{tmre64}.

We first solve the integrals over `$q\,'$' (in particular solve 
the $q'$ integral in $A_m$, equation~(\ref{am}))
and `$\alpha$' in $\siga{a}{}$ (equation~(\ref{gm1})) using the {\it 
stationary phase} approximation \citep{lghthll2001}. 
For an oscillatory integral with rapidly changing phase, 
most of the contribution to the integral cancels due to destructive 
superposition of oscillatory functions. Therefore, the phase can be 
approximated by its Taylor expansion around the stationary phase 
point, that is the point at which phase change is zero. In addition, the 
non-oscillatory part of the integrand is simply replaced by its 
value at the stationary point.

We begin by writing $\siga{a}{}(R)$ as 
\begin{equation}
\siga{a}{}(R)  =  h(R)\exp\left[\rmi \int^R k(R'')\dmath R''\right]\,. 
\end{equation}
This essentially divides $\siga{a}{}(R)$ into a slowly varying function 
$h(R)$ of $R$ and a fast varying oscillatory function of $R$. Substituting 
this in equation~(\ref{am}) we get,
\begin{equation}
 A_m(\alpha) = \intgl \dmath q' R'^{3/2}\,h(R')\,\exp(\rmi\psi)\,,
\label{am_1}
\end{equation}
where $\psi = \int^R k(R'')\dmath R'' - \alpha q'$, is the phase 
of the oscillatory part of the integral over $q\,'$. Any point 
$R = R_*$ (or equivalently $q = q_*$) is called a stationary 
point if at $q = q_*$, $\dmath \psi/\dmath q = 0$, which when 
substituted for $\psi$ gives the condition, $R_*k(R_*) = \alpha$. 
Since the phase is nearly constant at the stationary point (which 
in turn gives the leading contribution to the integral), we shall replace 
$\psi(R)$ with its Taylor expansion around $R_*$,
\begin{equation*}
\psi(R) = \psi(R_*) + (q - q_*)^2 \varrho\,,
\end{equation*}
where, $\varrho = R_*^2(\dmath k/\dmath R|_{R_*} + \alpha/R_*^2)/2$. 
We retain terms up to second order in the Taylor expansion. Other 
parts of the integrand are replaced by their value at $R_*$ and 
can be taken out of the integral. All this put together in the 
expression of $A_m$ given in Eq.~(\ref{am_1}) gives,
\begin{align}
A_m(\alpha) =& R_*^{3/2}\siga{a}{}(R_*)\,\exp[-\rmi\alpha q_*]
\intgl \dmath q'\exp\left[\rmi(q' - q_*)^2 \varrho\right]\,,\nonumber\\
=& R_*^{3/2}\siga{a}{}(R_*)\,\exp[-\rmi\alpha q_*] 
\left[\sqrt{\frac{\pi}{|\varrho|}}\exp\left[\rmi\, {\rm sgn}(\varrho)\frac{\pi}{4}\right]\right]\,.
\end{align}
For the second equality above we have used the Gaussian 
integral,
\begin{equation}
\intgl \dmath x \exp\left[\pm\rmi \lambda x^2\right] = 
\sqrt{\frac{\pi}{|\lambda|}}\exp\left[\pm\rmi\, {\rm sgn}(\lambda)\frac{\pi}{4}\right]\,,
\label{gausian}
\end{equation}
and ${\rm sgn}(\lambda)$ is the sign of $\lambda$. We next use this 
value of $A_m(\alpha)$ in the expression of $\siga{a}{\rm dt}$ given 
in equation~(\ref{sigdt3}) and the fact that $\siga{a}{} 
= \siga{a}{\rm dt}$ (because the indirect term equals zero for the present 
formulation), to get
\begin{align}
\siga{a}{}(R) =& \,\frac{2G\siga{d}{}}{R^{5/2}\kappa^2}\sum_{n = 1}^{\infty}\left(\frac{n^2}{n^2 - s^2}\right)
\intgl\frac{\dmath\alpha}{2\pi}\,N(\alpha,m)\,R_*^{3/2} h(R_*)\,\rme^{-\chi}\,\times\nonumber\\
&\times \left[B_n\sqrt{\frac{\pi}{|\varrho|}}\exp\left(\rmi\, {\rm sgn}(\varrho)\frac{\pi}{4}\right)\right]
\exp[\rmi\varphi]\,,
\label{sigdt4}
\end{align}
where the phase $\varphi = \int^{R_*} k(R'')\dmath R''-\alpha (q_* - q)$. 
At the stationary phase point $\dmath \varphi/\dmath\alpha = 0$, 
which on substitution of $\varphi$, gives the stationary phase 
point as $q = q_*$ or $R = R_*$. Hence integral over $\alpha$ in 
above equation, on applying the stationary phase 
approximation, simplifies to
\begin{align}
\siga{a}{}(R) =& \,\frac{G\siga{d}{}}{\pi R^{5/2}\kappa^2}\sum_{n = 1}^{\infty}\left(\frac{n^2}{n^2 - s^2}\right)
N(kR,m)\,R^{3/2} \siga{a}{}(R)\,B_n(\alpha,\chi)\,\times\nonumber\\
&\times \sqrt{\frac{\pi}{|\varrho|}}\exp\left(\rmi\, {\rm sgn}(\varrho)\frac{\pi}{4}\right)
\intgl\dmath\alpha\,\,\rme^{-\rmi\alpha^2 \varsigma}\,,\nonumber\\
=& \,\frac{2G\siga{d}{}}{2R\kappa^2}\sum_{n = 1}^{\infty}\left(\frac{n^2}{n^2 - s^2}\right)
N(kR,m)\,\siga{a}{}(R)\,B_n(\alpha,\chi)\,\times\nonumber\\
&\times \sqrt{\frac{1}{|\varrho\varsigma|}}\exp\left[\rmi\left({\rm sgn}(\varrho)\frac{\pi}{4}
- {\rm sgn}(\varsigma)\frac{\pi}{4}\right)\right]\,.
\label{sigdt5}
\end{align}
Here $\varsigma$ is defined as $(1/2R_*)(\dmath R_*/\dmath\alpha)$ and 
equation~(\ref{gausian}) is used to write the second equality. It can be 
noted from the definitions of $\varrho$ and $\varsigma$ that: 
\begin{enumerate}
\item $\varrho\,\varsigma = \frac{1}{4}$\,, and
\item ${\rm sgn}(\varrho) = {\rm sgn}(\varsigma)$\,. 
\end{enumerate}
\nin
Both these relation are obtained using $\alpha = 
R_*k(R_*)$, which was derived earlier. The standard WKB approximation 
is $|kR| \gg m$. Also $kR = \alpha$ at the 
stationary phase point. This gives an equivalent condition for 
WKB as $|\alpha| \gg m$. The asymptotic form of $N(\alpha,m)$ 
for $\alpha \gg m$ is 
\begin{equation}
N(\alpha,m) \sim \frac{2\pi}{(\alpha^2 + m^2)^{1/2}} \sim \frac{2\pi}{|\alpha|}\,. 
\label{NM_aysmtotic}
\end{equation}
\nin 
All these put together reduce equation~(\ref{sigdt5}) to 
\begin{align}
\siga{a}{}(R) =& \frac{2\pi G\sigd{}|k|}{\kappa^2}\sum_{n = 1}^{\infty}\left(\frac{n^2}{n^2 - s^2}\right)
\frac{2}{\chi}\rme^{-\chi}I_n(\chi)\siga{a}{}(R)\,,
\label{sigdt6}
\end{align}
which implies
\begin{align}
\kappa^2 - (\omega - m\Omega)^2 = 2\pi G\sigd{}|k|\mathcal{F}(s,\chi)\,,
\label{wkb}
\end{align}
where 
\begin{equation}
\mathcal{F}(s,\chi) \,=\, \frac{2}{\chi}\,(1 - s^2)\, e^{-\chi} \,\sum_{n=1}^{\infty}\frac{I_{n}(\chi)}{1 - s^2/n^2}\,.
\label{F}
\end{equation}
\nin
Thus we find that in the local approximation our equation reduces to the standard WKB dispersion relation \citep{tmre64,bt08}.
 
\section*{Acknowledgements}
We would like to thank S. Sridhar for many useful discussions during the course of this work.

\bibliographystyle{mn2e}

\begin{thebibliography}{}
\bibitem[Anderson et al.(1999)]{lapack} Anderson et. al.\ 1999,
 LAPACK Users' Guide (3rd ed., Society for Industrial and Applied Mathematics) 
\bibitem[Binney \& Tremaine(2008)]{bt08} Binney, J., \& Tremaine, S.\ 2008, Galactic 
Dynamics (2ed., Princeton: Princeton University Press)
\bibitem[Clampin et al.(2003)]{clamp03} Clampin, M., Krist, 
J.~E., Ardila, D.~R., et al.\ 2003, \aj, 126, 385 
\bibitem[Goldreich 
\& Tremaine(1979)]{gold79} Goldreich, P., \& Tremaine, S.\ 1979, \apj, 233, 857 
\bibitem[Gulati et al.(2012)]{gss12} Gulati, M., Saini, 
T.~D., \& Sridhar, S.\ 2012, \mnras, 424, 348 
\bibitem[Heap et al.(2000)]{heap00} Heap, S.~R., Lindler, 
D.~J., Lanz, T.~M., et al.\ 2000, \apj, 539, 435 
\bibitem[Jalali 
\& Tremaine(2012)]{jt12} Jalali, M.~A., \& Tremaine, S.\ 2012, \mnras, 421, 2368 
\bibitem[Kalnajs(1971)]{klnjs71} Kalnajs, A.~J.\ 1971, \apj, 
166, 275 
\bibitem[Lauer et al.(1993)]{lau93} Lauer, T.~R., Faber, 
S.~M., Groth, E.~J., et al.\ 1993, \aj, 106, 1436 
\bibitem[Lauer et al.(1996)]{lau96} Lauer, T.~R., Tremaine, 
S., Ajhar, E.~A., et al.\ 1996, \apjl, 471, L79 
\bibitem[Lighthill(2001)]{lghthll2001} Lighthill, J.\ 2001, Waves 
in Fluids, by James Lighthill, pp.~520.~ISBN 0521010454.~Cambridge, UK: 
Cambridge University Press, December 2001.,  
\bibitem[Marsh et al.(2006)]{marsh06} Marsh, K.~A., Dowell, 
C.~D., Velusamy, T., Grogan, K., \& Beichman, C.~A.\ 2006, \apjl, 646, L77 
\bibitem[Press et al.(1992)]{prs92} Press, W.~H., Teukolsky, S.~A., Vetterling, W.~T., \& Flannery, B.~P.\ 1992, Numerical Recipes (2nd ed.,
Cambridge: University Press)
\bibitem[Reichard et al.(2009)]{richrd09} Reichard, T.~A., 
Heckman, T.~M., Rudnick, G., et al.\ 2009, \apj, 691, 1005 
\bibitem[Sridhar \& Saini (2010)]{st10} Sridhar, S., \& Saini, T.D.\ 2010, \mnras, 404, 527
\bibitem[Telesco et al.(2000)]{tel00} Telesco, C.~M., Fisher, 
R.~S., Pi{\~n}a, R.~K., et al.\ 2000, \apj, 530, 329 
\bibitem[Toomre(1963)]{tmre63} Toomre, A.\ 1963, \apj, 138, 
385 
\bibitem[Toomre(1964)]{tmre64} Toomre, A.\ 1964, \apj, 139, 
1217 
\bibitem[Tremaine(2001)]{tre01} Tremaine, S.\ 2001, \aj, 121, 1776
\end{thebibliography}

\end{document}